\documentclass[12pt]{article}

\usepackage{scicite}

\usepackage{times}

\usepackage{graphicx}

\def \aap{Astron.~Astrophys.}
\def \apjl{Astrophys.~J.}
\def \apj{Astrophys.~J.}
\def \asp{Astron.~Soc.~Pac.}
\def \epjd{Euro.~Phys.~J.~D}
\def \jcp{J.~Chem.~Phys.}
\def \jqsrt{J.~Quant.~Spectrosc.~Radiat.~Transfer}
\def \lnp{Lecture~Notes~Phys.}
\def \mnras{Mon.~Not.~Roy.~Astron.~Soc.}
\def \prl{Phys.~Rev.~Lett.}
\def \pr{Phys.~Rev.}
\def \sci{Science}

\newcommand{\lsim}{\,\rlap{\raise 0.35ex\hbox{$<$}}{\lower 0.7ex\hbox{$\sim$}}\,}
\newcommand{\gsim}{\,\rlap{\raise 0.35ex\hbox{$>$}}{\lower 0.7ex\hbox{$\sim$}}\,}

\newenvironment{sciabstract}{%
\begin{quote} \bf}
{\end{quote}}

\newcounter{lastnote}

\title{Strong Limit on a Variable Proton-to-Electron Mass Ratio from
  Molecules in the Distant Universe}

\author
{Michael T.~Murphy,$^{1\ast}$ Victor V.~Flambaum,$^{2}$ S\'{e}bastien Muller,$^{3}$\\
Christian Henkel$^{4}$\\
\\
\normalsize{$^{1}$Centre for Astrophysics and Supercomputing, Swinburne University of Technology,}\\
\normalsize{Mail H39, PO Box 218, Victoria 3122, Australia}\\
\normalsize{$^{2}$School of Physics, University of New South Wales, Sydney, N.S.W. 2052, Australia}\\
\normalsize{$^{3}$Academia Sinica Institute of Astronomy and Astrophysics,}\\
\normalsize{PO Box 23-141, Taipei, 106 Taiwan}\\
\normalsize{$^{4}$Max-Planck-Institut f\"{u}r Radioastronomie, Auf dem H\"{u}gel 69, 53121 Bonn, Germany}\\
\\
\normalsize{$^\ast$To whom correspondence should be addressed; E-mail:  mmurphy@swin.edu.au.}
}

\date{}

\begin{document} 



\maketitle


\begin{sciabstract}
  The Standard Model of particle physics assumes that the so-called
  fundamental constants are universal and unchanging. Absorption lines
  arising in molecular clouds along quasar sightlines offer a precise
  test for variations in the proton-to-electron mass ratio,
  \boldmath{$\mu$}, over cosmological time and distance scales. The
  inversion transitions of ammonia are particularly sensitive to
  \boldmath{$\mu$} compared to molecular rotational transitions.
  Comparing the available ammonia spectra observed towards the quasar
  B0218$+$357 with new, high-quality rotational spectra, we present
  the first detailed measurement of \boldmath{$\mu$} with this
  technique, limiting relative deviations from the laboratory value to
  \boldmath{$\left|\Delta\mu/\mu\right| < 1.8\times10^{-6}$} (95\%
  confidence level) at approximately half the Universe's current age
  -- the strongest astrophysical constraint to date.  Higher-quality
  ammonia observations will reduce both the statistical and systematic
  uncertainties in these measurements.
\end{sciabstract}

The Standard Model of particle physics assumes that the fundamental
constants of nature (or, at least, their low-energy limits) are the
same everywhere and at every epoch in the Universe. However, it cannot
itself justify this assumption, nor can it predict their values. Our
confidence in their constancy stems from Earth-bound laboratory
experiments conducted over human time-scales. Extrapolating to the
entire Universe seems unwise, especially considering that the physics
driving the Universe's accelerating expansion, labeled `Dark Energy',
is completely unknown. Nevertheless, the Earth-bound experiments
achieve impressive precision: time-variations in the fine-structure
constant, $\alpha\equiv e^2/\hbar c$, which measures the strength of
electromagnetism, are limited to
$\dot{\alpha}/\alpha=(-1.6\pm2.3)\times10^{-17}{\rm \,yr}^{-1}$
\cite{RosenbandT_08a}, while those in the proton-to-electron mass
ratio, $\mu\equiv m_{\rm p}/m_{\rm e}$ -- effectively the ratio of the
strong and electro-weak scales -- are limited to
$\dot{\mu}/\mu=(+1.6\pm1.7)\times10^{-15}{\rm \,yr}^{-1}$
\cite{BlattS_08a}. Still, more dramatic variations might have occurred
over the 13--14\,Gyr history of the Universe and the residual
variations in our small space-time region might remain undetectably
small. It is therefore imperative to measure the constants over
cosmological time- and distance-scales.

Variations in $\mu$ and/or $\alpha$ would manifest themselves as
shifts in the transition energies of atoms and molecules. By comparing
transition energies registered in spectra of astronomical objects with
laboratory values, possible variations can, in principle, be probed
over our entire observable Universe and through most of its history.
Due to the narrowness of the spectral features involved, absorption
lines arising in gas clouds along lines of sight to background quasars
are currently our most precise cosmological probes. For example, by
comparing various heavy element electromagnetic resonance transitions
in optical quasar spectra, 5-$\sigma$ evidence has emerged for
variations in $\alpha$ of $\sim$5 parts in $10^{6}$ at redshifts
$0.2<z<4.2$ \cite{WebbJ_99a,MurphyM_01a,MurphyM_03a,MurphyM_04a}.
Although a more recent statistical sample found no variation
\cite{ChandH_04a}, errors in the analysis prevent reliable
interpretation of those results \cite{MurphyM_07f,MurphyM_08a},
leaving open the possibility of a varying $\alpha$.

Similarly, tentative (3-$\sigma$) evidence for a fractional variation
in $\mu$ of $\sim$$20\times10^{-6}$ has recently come from 2 quasar
spectra containing many ultraviolet (UV) H$_2$ transitions at redshift
$z\sim2.8$ \cite{IvanchikA_05a,ReinholdE_06a}. Comparison of UV
heavy-element resonance lines and H{\sc \,i} 21-cm absorption is
sensitive to $\alpha^2\mu$ and, assuming $\alpha$ to be constant, has
yielded indirect null constraints on $\mu$-variation, albeit with
slightly worse precision than the direct H$_2$ method
\cite{TzanavarisP_07a}.

An alternative method for measuring $\mu$ at high redshift, suggested
recently by Flambaum \& Kozlov \cite{FlambaumV_07a}, is to utilize the
sensitivity to $\mu$-variation of the ammonia inversion transitions
\cite{VeldhovenJ_04a} near 24\,GHz. A shift in their frequencies due
to a varying $\mu$ can be discerned from the cosmological redshift by
comparing them to transitions with lower sensitivity to $\mu$. Good
candidates for comparison are rotational transitions of molecules such
as CO, HCO$^+$ and HCN because (i) their transition frequencies depend
mainly on $\mu$ and not on other fundamental quantities
(e.g.~$\alpha$); (ii) they are simple molecules, commonly detected in
the interstellar medium; (iii) their rest frequencies (80--200\,GHz)
are not vastly dissimilar to the NH$_3$ transitions' (cf.~comparison
of UV and 21-cm), thus reducing possible effects due to
frequency-dependent spatial structure in the background quasar's
emission. This final point is discussed in more detail below.

For rotational and NH$_3$ inversion transitions we may write the
apparent change in velocity or redshift of an absorption line due to a
variation in $\mu$ as
\begin{equation}
\frac{\Delta v}{c} = \frac{\Delta z}{1+z} = K_i \frac{\Delta\mu}{\mu}\,,
\end{equation}
where $K_i$ is the sensitivity of transition $i$ to $\mu$ and
$\Delta\mu/\mu\equiv(\mu_z-\mu_{\rm lab})/\mu_{\rm lab}$ for $\mu_{\rm
  lab}$ and $\mu_z$ the values of $\mu$ in the laboratory and
absorption cloud at redshift $z$, respectively. All rotational
transitions have $K_i=1$, so comparing them with each other provides
no constraint on $\Delta\mu/\mu$. However, the sensitivity of the
NH$_3$ inversion transitions is strongly enhanced, $K_i\approx4.46$
\cite{FlambaumV_07a}. That is, as $\mu$ varies the NH$_3$ transitions
shift relative to the rotational transitions.

Only one quasar absorption system displaying NH$_3$ absorption is
currently known, that at $z=0.68466$ towards quasar B0218$+$357. From
redshift uncertainty estimates for NH$_3$, CO, HCO$^+$ and HCN in the
literature, \cite{FlambaumV_07a} crudely estimated the precision
achievable to be $\delta(\Delta\mu/\mu)\approx2\times10^{-6}$.
However, \cite{FlambaumV_07a} cautioned that a proper measurement of
$\Delta\mu/\mu$ from NH$_3$ would require detailed, simultaneous fits
to all the molecular transitions and that significantly better
precision may be possible. Here we make the first detailed measurement
of $\mu$ using the NH$_3$ inversion transitions by comparison with
HCO$^+$ and HCN rotational transitions.

The only published NH$_3$ absorption spectra are those for the
($J$,$K$) = (1,1), (2,2) and (3,3) inversion transitions reported by
\cite{HenkelC_05a} towards B0218$+$357, reproduced in Fig.~1. The
channel spacing is $1.67{\rm \,km\,s}^{-1}$ for (1,1) \& (2,2) and
$3.3{\rm \,km\,s}^{-1}$ for (3,3). The spectra are normalized by a
low-order fit to the quasar continuum. See \cite{HenkelC_05a} for
observational and data reduction details. The signal-to-noise ratio
(SNR) for the flux is very high, $\sim$1000\,per channel, but since
$<$1\% of the continuum is absorbed, the effective SNR for the
optical depth is quite low.

HCO$^+$ and HCN(1--2) absorption towards B0218$+$357 was discovered
more than a decade ago \cite{WiklindT_95a}. New, high resolution
($\approx$$0.9{\rm \,km\,s}^{-1}$ per channel), high SNR
($\sim$100\,per channel) observations of these lines were recently
undertaken with the Plateau de Bure Interferometer, France; see
\cite{materials}.  Fig.~1 shows both spectra normalized by fits to the
quasar continuum.

All spectra were registered to the heliocentric reference frame;
possible errors in this procedure are discussed in \cite{materials} and
shown to be negligible.

Spectra representing the 1-$\sigma$ uncertainty in normalized flux per
channel were constructed for all the molecular spectra by calculating
the root-mean-square (RMS) flux variations in the continuum portions
of each transition. Since no large differences were observed either
side of the absorption for any transition, a simple constant error
model was adopted.

As equation 1 states, the signature of a varying $\mu$ would be a
velocity shift between the rotational and NH$_3$ inversion
transitions.  Complicating the measurement of any shift is the
`velocity structure' evident in Fig.~1: the profiles comprise
absorption from many gas clouds, all associated with the absorbing
galaxy but nevertheless moving at different velocities. The number and
velocity distribution of these `velocity components' are unknown and
must be determined from the data themselves. Each fitted velocity
component is represented by a Gaussian profile parametrized by its
optical depth, Doppler width and redshift. The best-fitting parameter
values are determined with a $\chi^2$-minimization code, {\sc vpfit}
\cite{vpfit}, designed specifically for fitting quasar absorption
lines. To determine the statistically preferred velocity structure,
the best fit (i.e.~minimized) values of $\chi^2$ per degree of
freedom, $\chi^2_\nu$, are compared for several fits with different
velocity structures. That with the lowest $\chi^2_\nu$ is taken as the
fiducial one (i.e.~an F-test to discriminate between models).

Measuring $\Delta\mu/\mu$ requires the assumption that the velocity
structure is the same in all transitions. This does {\it not} mean
that the ratio of optical depths of corresponding velocity components
in different transitions must be constant across the profile. Rather,
it means that the number and velocity distribution of components are
assumed to be the same in different transitions. We discuss this
assumption below but, since it must be made eventually, in practice
the velocity structure was determined by tieing together the redshifts
of corresponding velocity components in different transitions. The
high SNR rotational spectra clearly place the strongest constraints on
the velocity structure but the NH$_3$ spectra must be included to
measure $\Delta\mu/\mu$. The Doppler widths of corresponding
components were also tied together, effectively assuming a turbulent
broadening mechanism.

Figure 1 shows the fiducial 8-component fit. Note that the detailed
hyperfine structure of the HCN(1--2) and NH$_3$ transitions is
reflected in each velocity component. The relative hyperfine level
populations were fixed by assuming local thermodynamic equilibrium
(LTE). The laboratory data used in the fits are tabulated in
\cite{materials}. Given this fit, determining $\Delta\mu/\mu$ is
straightforward: a single additional free parameter is introduced for
the entire absorption system which shifts all the velocity components
of each transition according to its $K$-coefficient (equation 1). All
parameters in the fit, including the single value of $\Delta\mu/\mu$,
are varied by {\sc vpfit} to minimize $\chi^2$. The best-fit value is
$\Delta\mu/\mu=(+0.74\pm0.47)\times10^{-6}$, corresponding to a
(statistically insignificant) velocity shift between the NH$_3$ and
rotational transitions of $0.77\pm0.49$\,km\,s$^{-1}$. The 1-$\sigma$
error quoted here -- formed from the diagonal terms of the final
parameter covariance matrix -- derives entirely from the photon
statistics of the absorption spectra. A different (though intimately
related) approach to determining $\Delta\mu/\mu$ and its error is
discussed in \cite{materials}; it provides the same result.

Fitting too few velocity components causes large systematic errors in
such analyses \cite{MurphyM_08a}. A reliable measurement of
$\Delta\mu/\mu$ can only be derived from fits replicating all the
statistically significant structure in the absorption profiles.
Therefore, the fiducial velocity structure must be the statistically
preferred one. When fitting many transitions simultaneously, this may
be more complicated than the one preferred by human eye. The simplest
objective method to achieve this is demonstrated in Fig.~2 which shows
the decrease in $\chi^2$ as increasingly complex velocity structures
are fitted. When the fit is too simple to adequately describe the
data, quite different values of $\Delta\mu/\mu$ are found. On the
other hand, the 9-component `over-fitted' case provides a value and
error very similar to the fiducial 8-component model.

The consistency of the velocity structures in the two highest SNR
transitions, HCO$^+$ and HCN(1--2), was tested by fitting those
transitions independently. Again, different fits with different
velocity structures were compared to determine the statistically
preferred one. Both transitions are best fit by velocity structures
similar to that in Fig.~1 [see \cite{materials}], providing some
confidence that they can meaningfully be fitted simultaneously. These
independent velocity structures were applied to the NH$_3$ transitions
and new values of $\Delta\mu/\mu$ derived. When fitting only
HCO$^+$(1--2) and NH$_3$, $\Delta\mu/\mu=(+0.67\pm0.51)\times10^{-6}$;
for HCN(1--2) and NH$_3$, $\Delta\mu/\mu=(+0.88\pm0.51)\times10^{-6}$.
Neither value substantially deviates from our fiducial one. Note that
the marginal increase in the 1-$\sigma$ error when using a single
rotational transition indicates that the NH$_3$ spectra limit the
statistical uncertainty.

To consider potential systematic uncertainties, it is important to
recall our main assumption: the velocity components constituting the
absorption profiles have the same redshifts in different transitions.
While the HCO$^+$ and HCN(1--2) velocity structures are evidently
similar enough for measuring $\Delta\mu/\mu$, the NH$_3$ spectra have
too low SNR for a direct comparison. And since the observed
frequencies of the NH$_3$ ($\sim$14\,GHz) and rotational
($\sim$106\,GHz) transitions are somewhat different, it is possible
that, if the background source morphology is frequency dependent, some
NH$_3$ components might arise along slightly different sight-lines to
those components in the rotational profiles. This is considered
presently.

B0218$+$357 is a $z=0.944$ BL Lac object \cite{CohenJ_03a} lensed by a
nearly face-on Sa/Sab $z=0.68$ galaxy \cite{YorkT_05a} in which the
absorption occurs. Two lensed images, A and B, separated by 334\,mas,
straddle the lensing galaxy's center with image B much closer to the
center. An Einstein ring with diameter $\sim$300\,mas, centered near
image B, has also been identified \cite{PatnaikA_93a}. B0218$+$357
itself has a core--jet morphology with an unresolved
[$<$$1\times1$\,mas or $<$$7\times7$\,pc \cite{cosmology}]
flat-spectrum core dominating the observed 8.4\,GHz emission
\cite{BiggsA_03a}. The jet has a knotty structure extending over
$\sim$$10\times10$\,mas and, like other jets, is expected to have a
steep spectrum.

Various absorption lines have been detected in the $z=0.68466$
absorber, from H{\sc \,i}\,21-cm and OH below (rest-frame) 2\,GHz
\cite{CarilliC_93a,KanekarN_03b}, through six H$_2$CO transitions at
5--150\,GHz \cite{JethavaN_07a}, to H$_2$O at 557\,GHz
\cite{CombesF_97b} to name but a few. Furthermore, the molecular
absorption arises only towards image A
\cite{MentenK_96a,CarilliC_00a,MullerS_07a,JethavaN_07a}. The
flat-spectrum core should completely dominate at high frequencies;
that H$_2$CO and H$_2$O absorb most of the total high-frequency
continuum therefore implies that at least those transitions only arise
towards the core. All the observed molecular transitions have
consistent velocity structures (though most spectra have poorer
resolution and/or SNR than those studied here). Thus, the most
important velocity components in all transitions evidently arise
towards image A's flat-spectrum, compact core
\cite{HenkelC_05a,MullerS_07a}.

Nevertheless, indirect evidence suggests that the molecular clouds do
not completely cover the background source
\cite{WiklindT_99b,materials}. If the covering fraction is different
for the rotational and NH$_3$ transitions, some velocity components
may appear in one and not the other. Similar problems may arise
because some HCO$^+$ and HCN(1--2) velocity components may be
optically thick \cite{materials}. The spurious shifts in
$\Delta\mu/\mu$ these effects may cause are difficult to estimate in
general but in \cite{materials} we conduct several fits in which
different combinations of NH$_3$ velocity components are removed,
providing an estimate of $\pm0.7\times10^{-6}$. Another potential
systematic error is our assumption of LTE for the HCN(1--2) and NH$_3$
hyperfine structure populations. As mentioned above, removing
HCN(1--2) from the analysis barely changes the measured
$\Delta\mu/\mu$. Removing different parts of the hyperfine structure
from the NH$_3$ transitions results in maximum deviations of
$\pm0.3\times10^{-6}$ from our fiducial value of $\Delta\mu/\mu$
\cite{materials}.

Combining these two potential systematic errors in quadrature, we
obtain $\Delta\mu/\mu = (+0.74\pm0.47_{\rm stat}\pm0.76_{\rm
  sys})\times10^{-6}$, providing no evidence for cosmological
variations in $\mu$. The NH$_3$ spectra currently set both the
statistical and systematic errors. While, clearly, the SNR and
resolution directly determine the former, they also indirectly
influence the latter: the velocity structure of higher quality NH$_3$
spectra could be more directly compared with the rotational profiles
and limits on non-LTE hyperfine structure anomalies could be
constrained by the data themselves. That is, with improved NH$_3$
spectra, both the statistical and systematic error components can be
improved. Nevertheless, until the NH$_3$ data are improved, our final
result is a 2-$\sigma$ limit on variation in $\mu$ from this single
absorber: $\left|\Delta\mu/\mu\right| < 1.8\times10^{-6}$. This
corresponds to $<1.9$\,km\,s$^{-1}$ shift between the NH$_3$ and
rotational transitions.

Our new value of $\Delta\mu/\mu$ seems inconsistent with the current
tentative evidence for $\mu$-variation from H$_2$ absorption at
$z\sim2.8$, $\Delta\mu/\mu=(+24.4\pm5.9)\times10^{-6}$
\cite{ReinholdE_06a}.  However, reliable comparison is difficult
because cosmological time- and/or space-variations in $\mu$ remain
poorly constrained. Clearly, a statistical sample from both
techniques, covering a wide redshift range and with detailed
assessment of systematic effects, is highly desirable.

Assuming $\mu$ varies linearly with time, our measurement corresponds
to a drift of $\dot{\mu}/\mu=(-1.2\pm0.8_{\rm stat}\pm1.2_{\rm
  sys})\times10^{-16}{\rm \,yr}^{-1}$. However, this assumption is
only a convenient means for comparison with current limits from
laboratory atomic clocks,
$\dot{\mu}/\mu=(+1.6\pm1.7)\times10^{-15}{\rm \,yr}^{-1}$
\cite{BlattS_08a}; it is not motivated by any physical (necessarily
untested) varying-$\mu$ theory.

With so few measurements of $\mu$ distributed throughout the Universe,
each new measurement is an invaluable test of the most basic -- and
theoretically unjustifiable -- assumptions in the Standard Model: that
the laws of physics are universal and unchanging. The precision
demonstrated here highlights the importance of discovering many more
molecular absorbers to our knowledge of fundamental physics.


\bigskip

\noindent{\bf Supporting Online Material}\\
www.sciencemag.org\\
Materials and Methods\\
Figs.~S1 to S3\\
Tables S1 and S2\\
References and Notes

\bigskip

\noindent{11 February 2008; accepted 22 May 2008}

\clearpage

\begin{figure}
\centerline{\includegraphics[width=0.95\columnwidth]{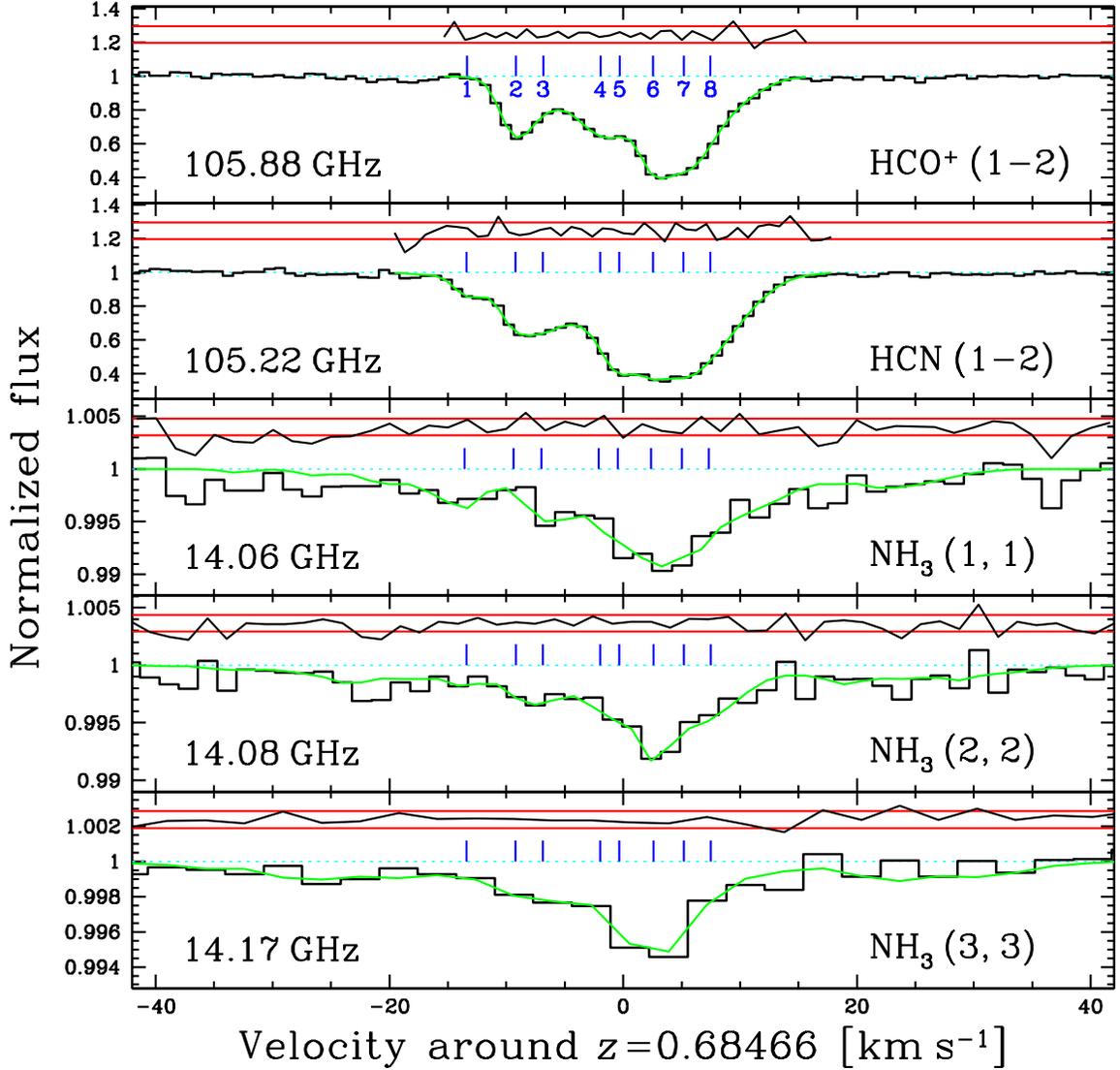}}
\caption{Spectra of the molecular transitions used in this study
  registered to a heliocentric velocity scale centered on $z=0.68466$.
  The nominal observed frequencies are noted in each panel.  The data,
  normalized by fits to their continua, are plotted as black
  histograms. Tick-marks above the spectra show the positions of
  velocity components in our fiducial 8-component fit (solid line
  following the data). Note that the HCN and NH$_3$ transitions have
  complex hyperfine structure reflected in each velocity component;
  the tick marks show the position of the strongest hyperfine
  component in LTE [see \cite{materials}]. Residuals between the fit
  and data, normalized by the (constant) error array, are plotted
  above the spectra, bracketed by horizontal lines representing the
  $\pm1\,\sigma$ level. The fit contains 57 free parameters: an
  optical depth for each component in each transition (5$\times$8
  parameters) plus a Doppler width and redshift for each component
  (8$+$8 parameters) and a single value of $\Delta\mu/\mu$. The fitted
  line parameters are tabulated in \cite{materials}.}
\end{figure}

\begin{figure}
\centerline{\includegraphics[width=0.95\columnwidth]{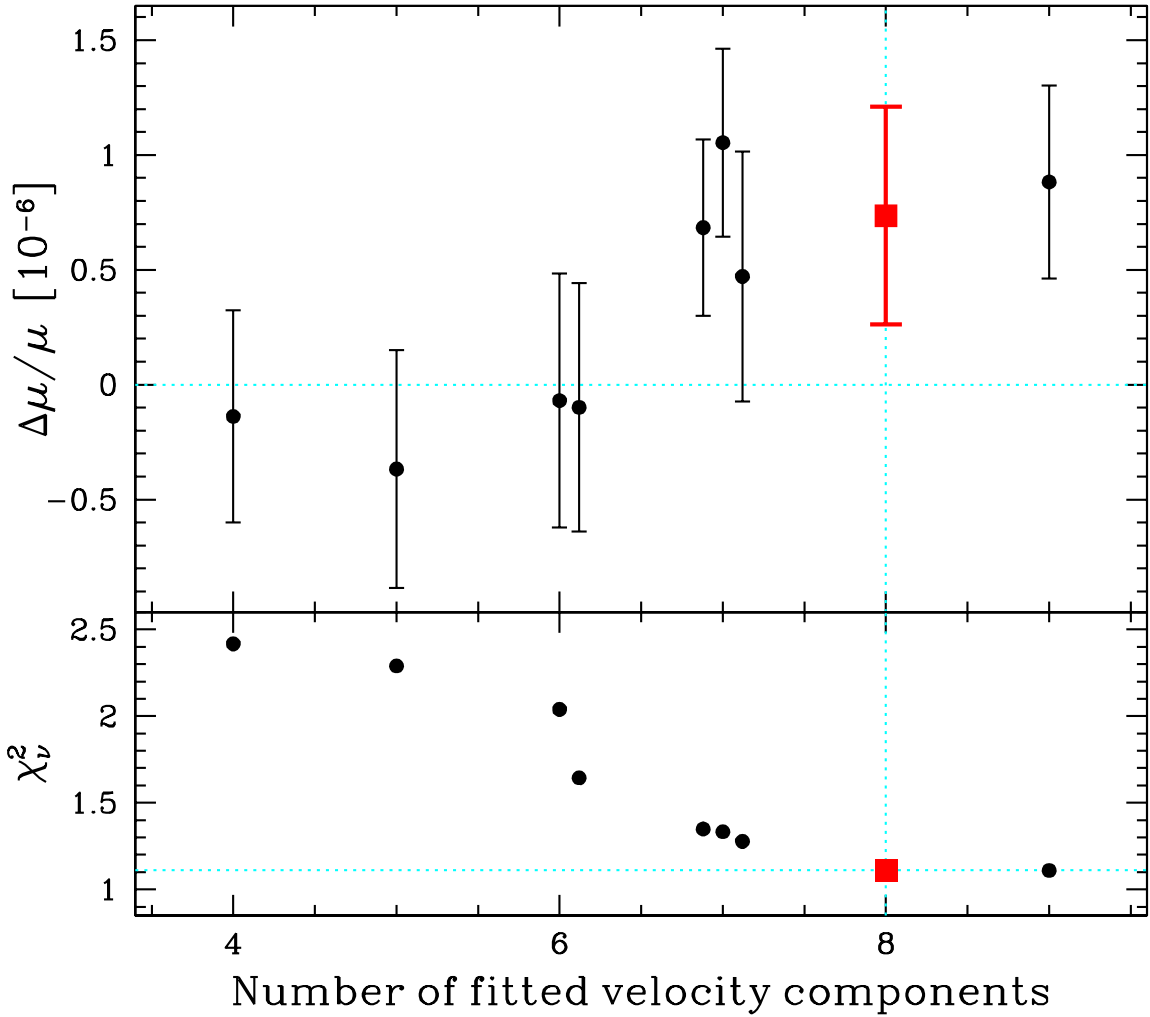}}
\caption{Variation in $\Delta\mu/\mu$ and $\chi^2$ per degree of
  freedom, $\chi^2_\nu$, of different velocity structures
  characterized by the number of fitted absorption components.
  $\chi^2_\nu$ is defined as $\chi^2/\nu\equiv\sum_j^{N_{\rm
      d}}[d_j-m(j)]/\sigma_j^2$ for $d_j$ the $j^{\rm th}$ data value
  with variance $\sigma_j^2$ and model value $m(j)$. The sum is over
  all $N_{\rm d}=223$ data points; $\nu\equiv N_{\rm d}-N_{\rm par}$
  for $N_{\rm par}$ free model parameters. Our fiducial 8-component
  ($N_{\rm par}=57$) result is highlighted with square points.
  Different components were added/removed to/from the fiducial fit to
  form each initial velocity structure and {\sc vpfit} was run again
  to minimize $\chi^2$ by varying all free parameters. Two different
  initial fits with 6 components and three fits with 7 components were
  possible; the different results are offset in the plot for clarity
  in these cases. Large $\chi^2_\nu$ values for $\le6$ components
  indicate that those fits are not statistically acceptable. Of the
  remaining fits, the 8-component fit has the lowest $\chi^2_\nu$.
  Note that the 9-component fit has a smaller $\chi^2$ (because more
  parameters are being fitted) but a marginally higher $\chi^2_\nu$,
  indicating that it is less statistically preferred than the
  8-component fit. Only statistical error bars on $\Delta\mu/\mu$ are
  shown; see text for discussion about systematic errors.}
\end{figure}

\clearpage

{\Huge\bf SUPPORTING ONLINE MATERIAL}

\section*{Materials and methods}

\subsection*{New observations of HCO$^+$ and HCN(1--2)}

The HCO$^+$ and HCN(1--2) spectra were obtained with the Plateau de
Bure Interferometer (PdBI) with the 3-mm receivers, the HCO$^+$(1--2)
line being redshifted to 105.882\,GHz and the HCN(1--2) line to
105.221\,GHz. Five or six antennas were used in compact
configurations, and the two images, A and B, of B0218$+$357 were not
resolved from each other. The continuum flux density of B0218$+$357
was estimated to be $\sim$0.5\,Jy.

The spectra presented here are averages of spectra taken on several
observing runs: 28 and 29 June 2005, 15 July 2006 and 23 March 2007
for the HCO$^+$ data; 5 July 2005, 19 August 2006 and 23 March 2007
for the HCN data. The integration time varied from 30-min to
$\sim$2-hr. Within the SNR of individual integrations, we do not see
time variation in the absorption profiles between 2005 and 2007. Note
that in March 2007 the HCO$^+$ and HCN data were obtained
simultaneously with dual polarization receivers.

The frequency dependence of the instrument across the spectral
bandwidth (80\,MHz with 256 channels of width $\sim$0.3\,MHz) was
calibrated by observing a strong radio quasar for several minutes.
Its amplitude and phase as a function of radio frequency were solved
for by fitting high order polynomials (10 and 20, respectively). The
RMS residuals of the bandpass calibration were typically less than
1--2\%.  The visibilities were self-calibrated with respect to the
continuum emission of B0218$+$357 itself: the phase was referenced to
the barycenter of the continuum emission and the amplitude was
normalized to the total continuum flux. The spectra observed with the
different baselines of the interferometer and at different epochs were
then co-added. Third-degree baselines (derived on channels free of
absorption) were subtracted from these final spectra to remove small
residuals in the radio bandpass calibration, instrument drifts and/or
atmospheric fluctuations.

\subsection*{Laboratory molecular data}

Table S1 provides all the laboratory molecular data used in our
analysis.  Laboratory frequencies for HCO$^+$ and HCN(1--2) were taken
from \cite{DeLuciaF_69a} and \cite{SastryK_81a} respectively, as
recommended by \cite{PickettH_98a}. Frequencies for the NH$_3$
inversion lines were taken from \cite{KukolichS_67a}. Including the
hyperfine structure of the HCN(1--2) and NH$_3$ inversion transitions
when fitting the absorption spectra is very important. The relative
optical depths of the HCN(1--2) hyperfine components, assumed to be in
LTE, were taken from \cite{PickettH_98a} while those for the NH$_3$
lines were computed using the formula presented by
\cite{KukolichS_67a}.

\subsection*{Alternative \boldmath{$\chi^2$} minimization scheme}

The simplest, fastest and most robust method of determining
$\Delta\mu/\mu$ from a multi-component, simultaneous fit to different
transitions is simply to include $\Delta\mu/\mu$ as an additional free
parameter in the $\chi^2$ minimization process. This was our approach
in the main text. An alternative approach is to conduct many
simultaneous fits to the absorption spectra, each with a fixed input
value of $\Delta\mu/\mu$ but allowing all other free parameters to vary,
to build up a `$\chi^2$ curve' -- a plot of $\chi^2$ versus
$\Delta\mu/\mu$. The best-fitting value of $\Delta\mu/\mu$ is the one
which produces the lowest value of $\chi^2$, $\chi^2_{\rm min}$. The
functional form of $\chi^2$ implies that the $\chi^2$ curve should be
near parabolic and smooth on scales similar to the 1-$\sigma$ error on
$\Delta\mu/\mu$.  The error can therefore be determined from the width
of the $\chi^2$ curve near the minimum; the curve's smoothness is
obviously essential to this procedure -- see
\cite{MurphyM_07f,MurphyM_08a}, cf.~\cite{ChandH_04a}.

Figure S1 shows the $\chi^2$ curve for our fiducial 8-component fit to
the rotational and NH$_3$ inversion spectra. For convenience, we plot
$\Delta\chi^2\equiv\chi^2-\chi^2_{\rm min}$ normalized by the lowest
value of $\chi^2_\nu$ -- i.e.~$\chi^2$ per degree of freedom -- in the
curve, $\chi^2_{\nu{\rm ,min}}$. This representation allows the
1-$\sigma$ error in $\Delta\mu/\mu$ to be determined directly from the
width of the curve at $\Delta\chi^2/\chi^2_{\nu{\rm ,min}}=1$. The
result from this process is
$\Delta\mu/\mu=(+0.75\pm0.45)\times10^{-6}$, in very close agreement
with the value determined simply by fitting $\Delta\mu/\mu$ as an
additional free parameter,
$\Delta\mu/\mu=(+0.74\pm0.47)\times10^{-6}$.

\subsection*{Errors from bandpass calibration and baseline stability?}

Possible observational systematic uncertainties include bandpass
calibration errors and variable spectral baselines. The former may
affect the PdBI (i.e.~iterferometric) rotational spectra while the
latter will be most severe for the single-dish NH$_3$ spectra which
were obtain with the Effelsberg 100-m Ratio Telescope in Germany. Both
these effects could potentially lead to small ripples in the
absorption spectra. As mentioned above, the RMS bandpass calibration
errors for the new rotational spectra are $<$2\%.  Additionally, the
fact that $\Delta\mu/\mu$ is insensitive to the rotational transition
used in the analysis suggests that any residual ripples in the HCO$^+$
or HCN(1--2) spectra are unimportant. The poorer SNR (on the optical
depth) of the NH$_3$ transitions makes individually removing them a
less stringent test.  Nevertheless, removing NH$_3$\,(1,1) yields
$\Delta\mu/\mu=(+0.67\pm0.62)$, removing NH$_3$\,(2,2) gives
$(+0.65\pm0.62)$ and eliminating NH$_3$\,(3,3) provides
$(+0.84\pm0.49)$, all in units of $10^{-6}$. Wald--Wolfowitz runs
tests on the unabsorbed portions of the NH$_3$ spectra also reveal no
evidence for significant ripples. Therefore, there is no evidence for
significant systematic effects due to bandpass calibration and/or
baseline stability problems.

\subsection*{Alternative velocity structures}

As described in the main text, we fitted several different velocity
structures in our search for the one with the lowest $\chi^2_\nu$,
i.e.~the fiducial 8-component model plotted in Fig.~1. For each
velocity structure, {\sc vpfit} was run again to minimize $\chi^2$ by
varying all free parameters of the fit. These alternative fits
provided the black points in Fig.~2. For comparison, Fig.~S2 plots
these alternative structures in the same way as Fig.~1. By inspecting
the residual spectra (plotted above each transition), it is easy to
see why the fits with fewer than 8 components result in larger values
of $\chi^2_\nu$ compared to the fiducial model. The residuals take
many-pixel excursions outside the 1-$\sigma$ ranges at similar
velocities in different transitions -- a sure sign that real structure
exists in the absorption profiles which is left unaccounted for by the
fit. For example, the human eye readily picks out 5 or 6 components in
the rotational profiles but the two different 6-component models
produce very poor fits (two right-hand panels in middle row) with
$\chi^2_\nu> 1.6$ in Fig.~2. On the other hand, the 9-component model
provides ``too good'' a fit, as demonstrated by its marginally {\it
  higher} $\chi^2_\nu$ value in Fig.~2 compared to the fiducial model.
The best-fit line parameters for the fiducial 8-component model are
given in Table S2.

The high SNR of the HCO$^+$ and HCN(1--2) spectra implies that our
determination of the velocity structure is driven by those transitions
and not the NH$_3$ inversion lines. A good cross-check on the central
assumption in our analysis -- that the velocity structure is the same
in different transitions -- can therefore be made by fitting the
HCO$^+$ and HCN(1--2) spectra separately. Fig.~S3 compares our
independent fits of these rotational lines. Only the velocity
structures providing the lowest $\chi^2_\nu$ values are plotted. Seven
components provide the lowest-$\chi^2_\nu$ fit for the HCO$^+$ data
while 8 components are required to fit the HCN spectrum. However, as
for our fiducial simultaneous fit in Fig.~1, the bluest component in
HCN is not statistically required by the HCO$^+$ data. That is, both
the HCN and HCO$^+$ data can be individually fit by the same number of
components as used in the simultaneous fit. Also note the general
agreement between the velocities at which these components are best
fitted. In this sense the velocity structures of the two high SNR
rotational transitions are consistent.

\subsection*{Errors from morphology and variability of B0218$+$357?}

Quasars are not point sources at radio wavelengths but instead show
sometimes complex morphological structure. Indeed, each lensed image
of B0218$+$357 shows a core--jet morphology. If the absorbing
molecular clouds at $z=0.68466$ have smaller angular extent than the
background radio source, and the source's morphology varies between 24
and 178\,GHz (in the absorber's reference frame), then the NH$_3$ and
rotational transitions may arise along slightly different sight-lines.
This may, in principle, lead to systematic errors in the determination
of $\Delta\mu/\mu$.

As discussed in the main text, all the molecular absorption lines
studied here are very likely to arise only towards the core of image
A. Its angular extent, $<$$1\times1$\,mas, places an upper limit on
the spatial extent of the absorbing region of just $<$$7\times7$\,pc
\cite{HenkelC_05a,MullerS_07a,cosmology}. This is a typical size for
Galactic molecular clouds \cite{TerebeyS_86a}. Nevertheless, if the
core is close to the upper limit in angular extent and/or if the
absorbing clouds are smaller than $\sim$5\,pc, then it is possible
that the NH$_3$ and rotational profiles may have different velocity
structures. The similarity of the velocity structures of a variety of
transitions -- covering a larger frequency range than just those
studied here -- suggests this may be of little concern.

However, evidence for small-scale structure of the absorbing clouds
has been found in \cite{WiklindT_99b}. By examining the optical
obscuration and reddening of image A by dust expected to reside in the
absorbing clouds, \cite{WiklindT_99b} established that not all of
image A's optical emission is obscured and that its unobscured portion
is not even substantially reddened.  Therefore, assuming that the main
component of optical emission originates from image A's core (just as
for the radio emission), the absorbing clouds must not completely
cover the background emission. Since the emission region's size should
vary with frequency, the NH$_3$ profiles may contain a different
number of absorbing velocity components compared to the rotational
profiles (though some components may still be in common).

The flux variability of B0218$+$357 may also be of potential concern.
Again, if the absorbing clouds do not cover the core of image A and
its morphology changes with time, then observations at different
epochs may probe slightly different sight-lines. As detailed above,
the HCO$^+$ and HCN(1--2) data were obtained over several runs between
June 2005 and March 2007 with HCO$^+$ usually observed up to 1 month
before HCN. The NH$_3$ data were collected in August 2001 and June
2002.

It is difficult to use the available data to test specifically for
each of these effects. Nevertheless, the fact that the low SNR NH$_3$
profiles at least resemble the higher SNR rotational ones --
particularly the velocity of the peak optical depth -- allows a
general quantitative assessment of the likely effect on the
$\Delta\mu/\mu$ measurement. One test is to remove specific (groups
of) velocity components from the fits to the NH$_3$ spectra while
keeping the fiducial 8-component model for the rotational lines. Upon
removing a single velocity component -- that is, fitting 7 components
to the NH$_3$ spectra -- the different possible fits yield an RMS
deviation of $0.5\times10^{-6}$ from our fiducial $\Delta\mu/\mu$.
This is dominated by the results of removing either component 3 or
component 6 (see Fig.~1), which give deviations of
$\approx0.7\times10^{-6}$. Removing selected pairs of components
(e.g.~1 \& 3 or 5 \& 7 etc.) or selected combinations of three
components yield similar RMS deviations of $\approx0.5\times10^{-6}$;
these are again dominated by cases where either components 3 or 6 are
removed. Another test is to fit the NH$_3$ profiles with as few
components as is statistically acceptable (i.e.~$\chi^2_\nu\sim1$) --
in our case just two, components 3 \& 6 -- while keeping the full
velocity structure in the rotational lines. For this test we relaxed
the constraint that the widths of the rotational and NH$_3$ components
should be the same. We find a deviation of $0.45\times10^{-6}$ from
the fiducial value to $\Delta\mu/\mu=(+0.29\pm0.39)\times10^{-6}$.

Given these results, we adopt a representative systematic error on
$\Delta\mu/\mu$ of $\pm0.7\times10^{-6}$ due to possible differences
in the velocity structure of the rotational and NH$_3$ spectra. Note
that if higher resolution, higher SNR NH$_3$ spectra were obtained,
a more direct comparison of the NH$_3$ velocity structure with that of
the rotational transitions could be made, similar to the comparison of
the two rotational transitions made above. Therefore, high quality
NH$_3$ spectra should reduce the systematic error calculated here.

\subsection*{Optical depths}

Neither the PdBI rotational observations nor the Effelsberg NH$_3$
observations resolved the two lensed images from each other. Thus,
even though the molecular absorption is very likely to arise only
towards the core of image A, continuum flux from image B, the jet of
image A and possibly the Einstein ring will also contribute to the
spectra studied here.  The flat-spectrum lensed cores of images A and
B are thought to dominate the continuum flux beyond 14\,GHz and image
A contributes $>$70\% of the flux
\cite{PatnaikA_95a,HenkelC_05a,MullerS_07a}.  Therefore, the fraction
of image A's core continuum absorbed by the NH$_3$ and rotational
lines is about a factor of 1.4 larger than shown in Fig.~1.  Thus, if
some velocity components truly absorb all of image A's core continuum
(at the observed spectral resolution), they will appear flat-bottomed
but will not extend down to the zero flux level.

For simplicity, we did not attempt to take this potential effect into
account in the profile fits in Fig.~1. It is clearly not a problem for
the NH$_3$ inversion lines since they only absorb $\sim$1\% of the
total continuum flux. However, it is possible that some velocity
components in HCO$^+$ and HCN(1--2) are near saturation (see below),
even though we do not observe flat-bottomed profiles. We tested the
effect of this scenario on $\Delta\mu/\mu$ in the following way: 30\%
of the flux was subtracted from the rotational spectra and the
continuum levels were appropriately renormalized. Instead of Gaussian
absorption components we fitted Voigt profiles to allow saturation
effects in the fit. We conducted several fits with a range of quantum
mechanical damping coefficients to characterize the onset of damping
wings in the Voigt profiles. All fits gave results negligibly
different to our fiducial result, indicating that, under the
assumption that our fitted velocity structure reflects the real
velocity distribution of absorbing clouds, saturation effects do not
affect our measurement of $\Delta\mu/\mu$.

It is likely that some of the stronger components in the HCO$^+$ and
HCN(1--2) profiles are optically thick. Precise optical depths are
difficult to determine because the fraction of the background source
covered by the absorption clouds is also unknown. Wiklind \& Combes
\cite{WiklindT_95a} found optical depths of $\gsim1.9$ for HCN(1--2)
and $\gsim1.1$ for HCO$^+$(1--2). A similar conclusion was reached for
HCO$^+$(1--2) by \cite{MullerS_07a}. If several velocity components
with very similar velocities ($\Delta v\lsim2$\,km\,s$^{-1}$) and
different optical depths exist in the rotational spectra but are
fitted as a single component, then some spurious shifts with respect
to the optically thin NH$_3$ profile might be measured. The potential
systematic error on $\Delta\mu/\mu$ is therefore the same as that
caused by uncertainty in the NH$_3$ velocity structure, as quantified
above.

\subsection*{Errors from hyperfine structure?}

All the fits discussed in the main text and above assumed that the
relative optical depths of the HCN(1--2) and NH$_3$ hyperfine
structure components were those expected in LTE -- see Table S1.
However, non-LTE hyperfine structure effects have been observed in
HCN(0--1) emission lines from Galactic interstellar clouds
\cite{CernicharoJ_84a}. Nevertheless, if LTE conditions do not prevail
for the HCN(1--2) velocity components, then the effect on
$\Delta\mu/\mu$ seems to be small since removing HCN(1--2) from the
analysis does not change our measured $\Delta\mu/\mu$ substantially
(see main text). However, removing the NH$_3$ transitions is not
possible if we are still to measure $\Delta\mu/\mu$, nor are the
NH$_3$ spectra of high enough quality to directly determine the
hyperfine structure populations; the potential effect of non-LTE
hyperfine structure must be assessed differently.

NH$_3$ hyperfine structure `anomalies' have previously been found in
Galactic sources both in absorption \cite{MatsakisD_77a} and emission
\cite{StutzkiJ_84a}. The non-LTE conditions cause the $\Delta F_1\ne0$
`satellite' components with a given quadrupole quantum number, $F_1$,
to have different optical depth depending on the change in $F_1$
during the transition. Since complementary (i.e.~$F_1$=1--2 and
$F_1$=2--1) transitions straddle the main ($\Delta F_1=0$) hyperfine
components, non-LTE conditions cause asymmetries in the optical depth
profile of any given velocity component. A simple test to gauge the
maximum effect this may have on $\Delta\mu/\mu$ is to remove either
the high or low frequency satellite lines from the NH$_3$ fits. This
causes RMS deviations from our fiducial $\Delta\mu/\mu$ of
$0.3\times10^{-6}$. We adopt this as a second component to our
systematic error budget in the main text. Note that with higher
resolution, higher SNR NH$_3$ spectra it should be possible to conduct
consistency checks for strong non-LTE hyperfine structure effects,
thereby reducing the possible systematic error.

\subsection*{Reference frames}

All the spectra studied here were shifted to the heliocentric rest
frame so that any velocity shifts between them could be attributed to
a varying $\mu$. 

Depending on the format in which raw radio telescope data are
recorded, passed to the data reduction software and processed, one or
more small velocity shifts may be applied to them. These typically
convert between the local standard of rest (LSR) and heliocentric
frames. However, the Doppler correction for the radial velocity,
\begin{equation}
\frac{v_{\rm rad}}{c}=\frac{\nu_{\rm obs}^2-\nu_{\rm 0}^2}{\nu_{\rm obs}^2+\nu_{\rm 0}^2}\,,
\end{equation}
is approximated in many data reduction software packages (for
traditional reasons, not ones of simplicity or processing speed etc.)
by
\begin{equation}
\frac{v_{\rm rad}}{c}\approx\frac{\nu_{\rm obs}-\nu_{\rm 0}}{\nu_{\rm 0}}
\end{equation}
which applies when $v_{\rm rad}/c\ll 1$. Here $\nu_{\rm 0}$ is the
emitted frequency and $\nu_{\rm obs}$ is the observed frequency.
Since we are interested in small shifts between different transitions
with a precision of $\delta({\Delta\mu/\mu})\sim0.1\times10^{-6}$,
corresponding to a velocity shift of just $0.1{\rm \,km\,s}^{-1}$, we
must check whether the above approximation is accurate enough so as
not to cause systematic errors in $\Delta\mu/\mu$. The difference
between the above two expressions for the Doppler correction is
approximately $(v_{\rm rad}/c)^2$ for $v_{\rm rad}/c\ll 1$. So, for a
maximal heliocentric correction of $\sim$$30{\rm \,km\,s}^{-1}$, the
systematic error introduced by using the approximation is $\sim$$3{\rm
  \,m\,s}^{-1}$.  This corresponds to
$\delta({\Delta\mu/\mu})\sim3\times10^{-9}$, well below our current
statistical precision.

\clearpage

\begin{figure}
\renewcommand \thefigure{S1}
\centerline{\includegraphics[width=\textwidth]{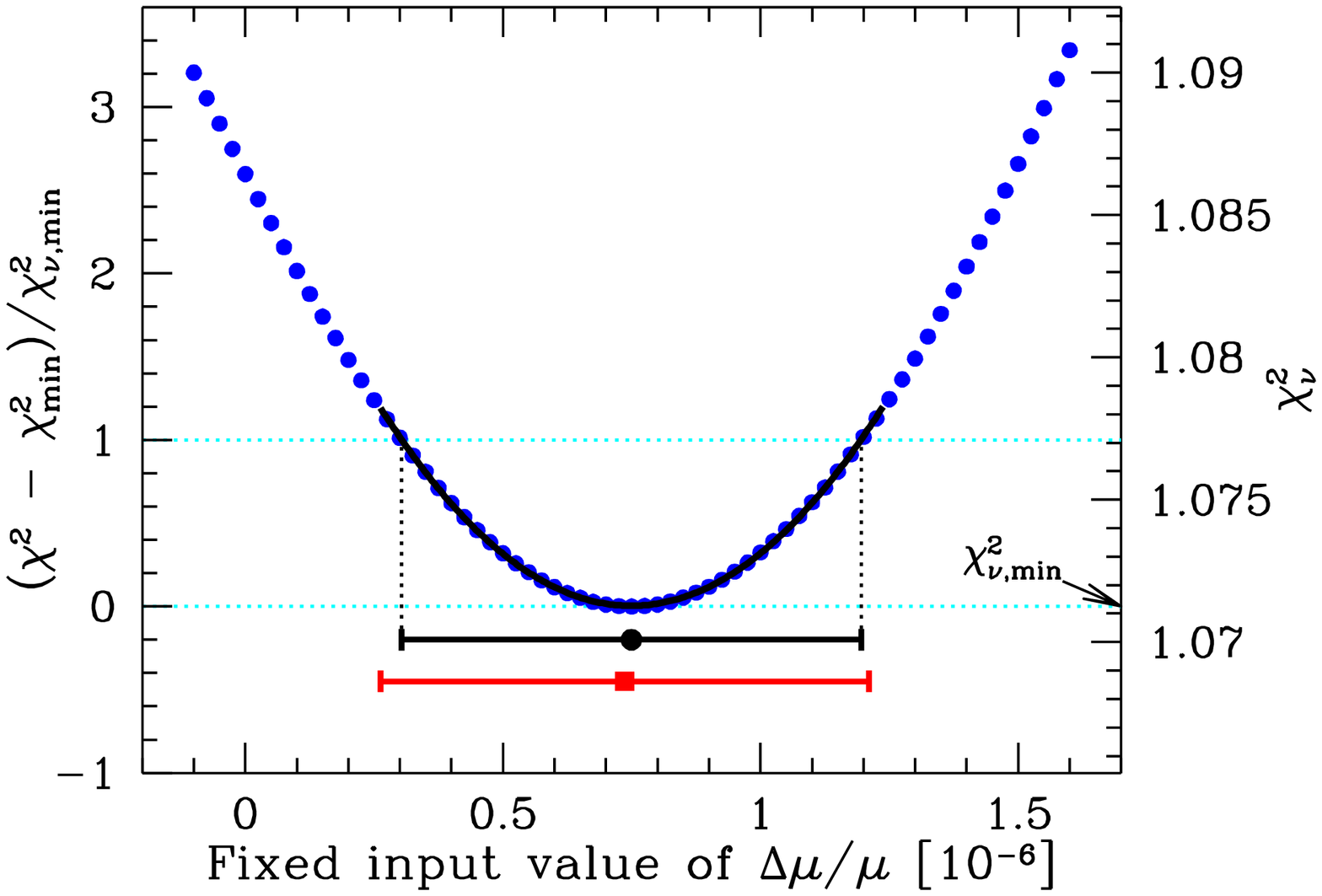}}
\caption{$\chi^2$ curve for the fiducial 8-component
model fit. Note its (required) smoothness and (expected)
near-parabolic shape. The right-hand vertical scale shows $\chi^2_\nu$
which takes a minimum value of $\chi^2_{\nu{\rm,min}}=1.07125$. The
left-hand vertical scale shows $\Delta\chi^2/\chi^2_{\nu{\rm ,min}}$,
thereby allowing the 1-$\sigma$ error in $\Delta\mu/\mu$ to be
immediately read off the graph (dotted black lines). In practice, we
measured the position at which $\chi^2$ takes its minimum and the
curve's width by fitting a parabola to the central points; the fit,
marked by the black solid line, demonstrates how close to parabolic
that part of the curve is. The black circle and error-bar represent
the result from this fit, $\Delta\mu/\mu=(+0.75\pm0.45)\times10^{-6}$,
which closely matches our fiducial value represented by the red/grey
square and error-bar.}
\end{figure}

\begin{figure*}
\renewcommand \thefigure{S2}
\centerline{
 \vbox{
  \hbox{
   \includegraphics[width=0.32\textwidth]{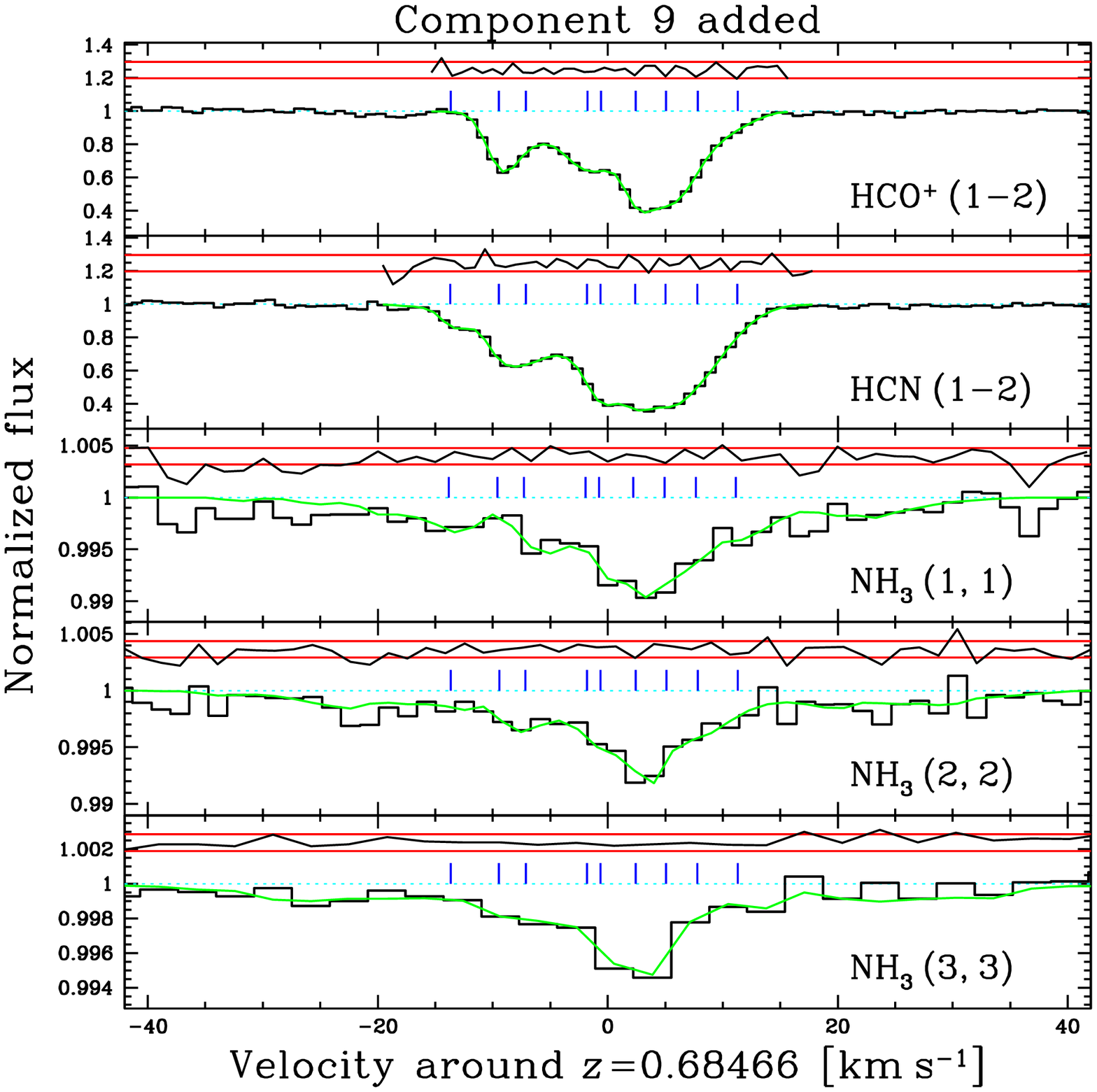}
   \hspace{0.01\textwidth}
   \includegraphics[width=0.32\textwidth]{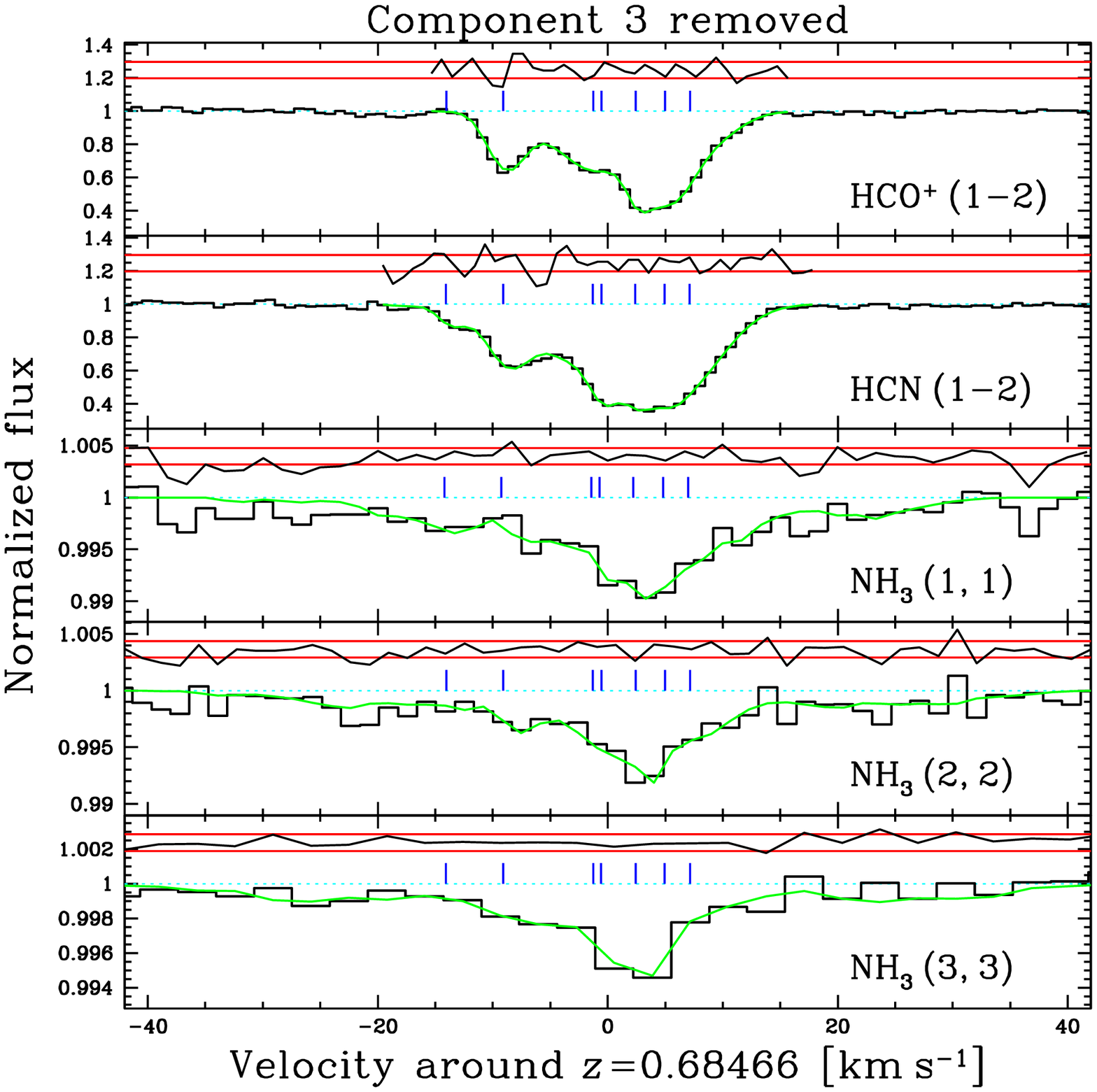}
   \hspace{0.01\textwidth}
   \includegraphics[width=0.32\textwidth]{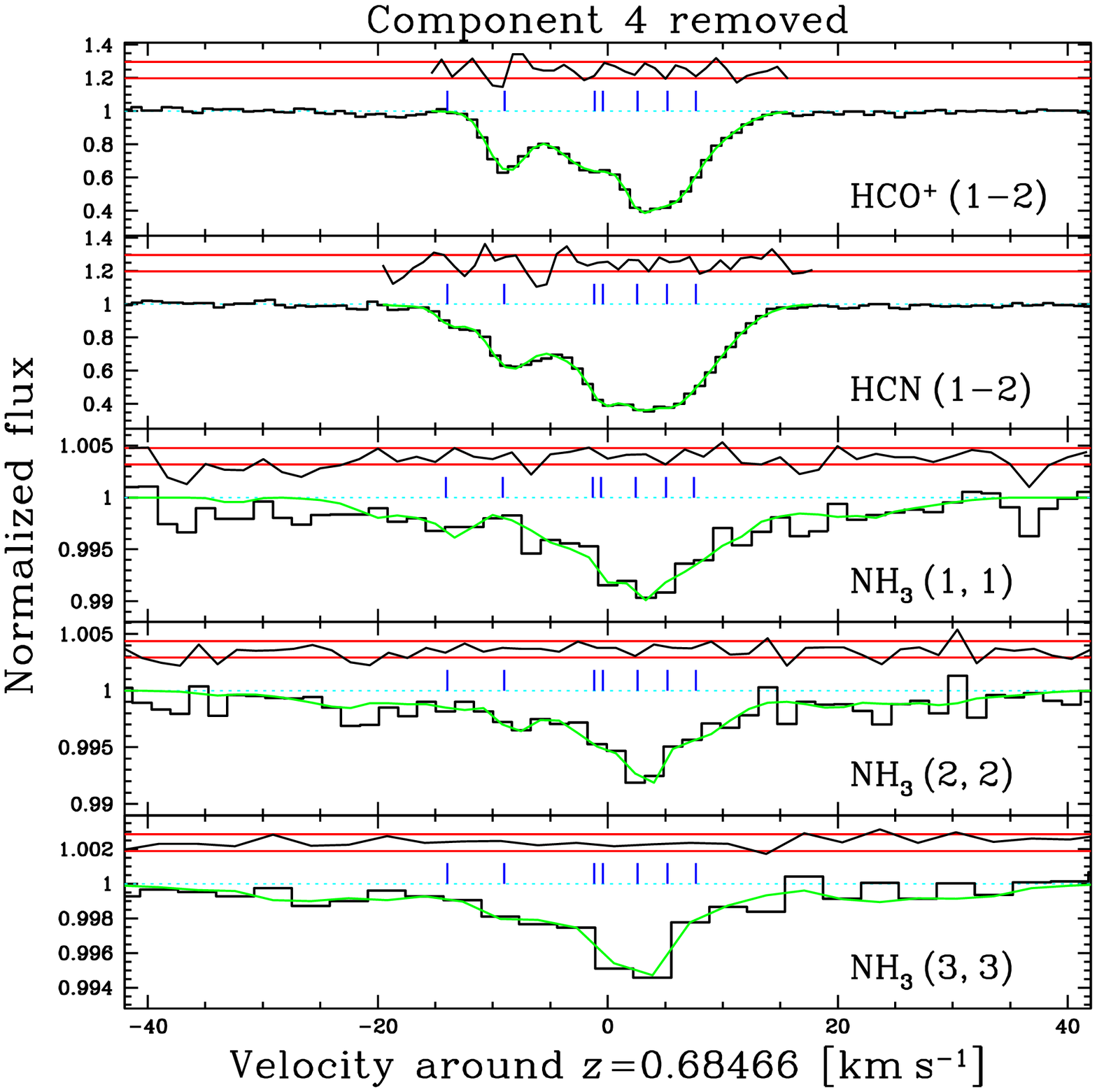}
  }
  \vspace{0.01\textheight}
  \hbox{
   \includegraphics[width=0.32\textwidth]{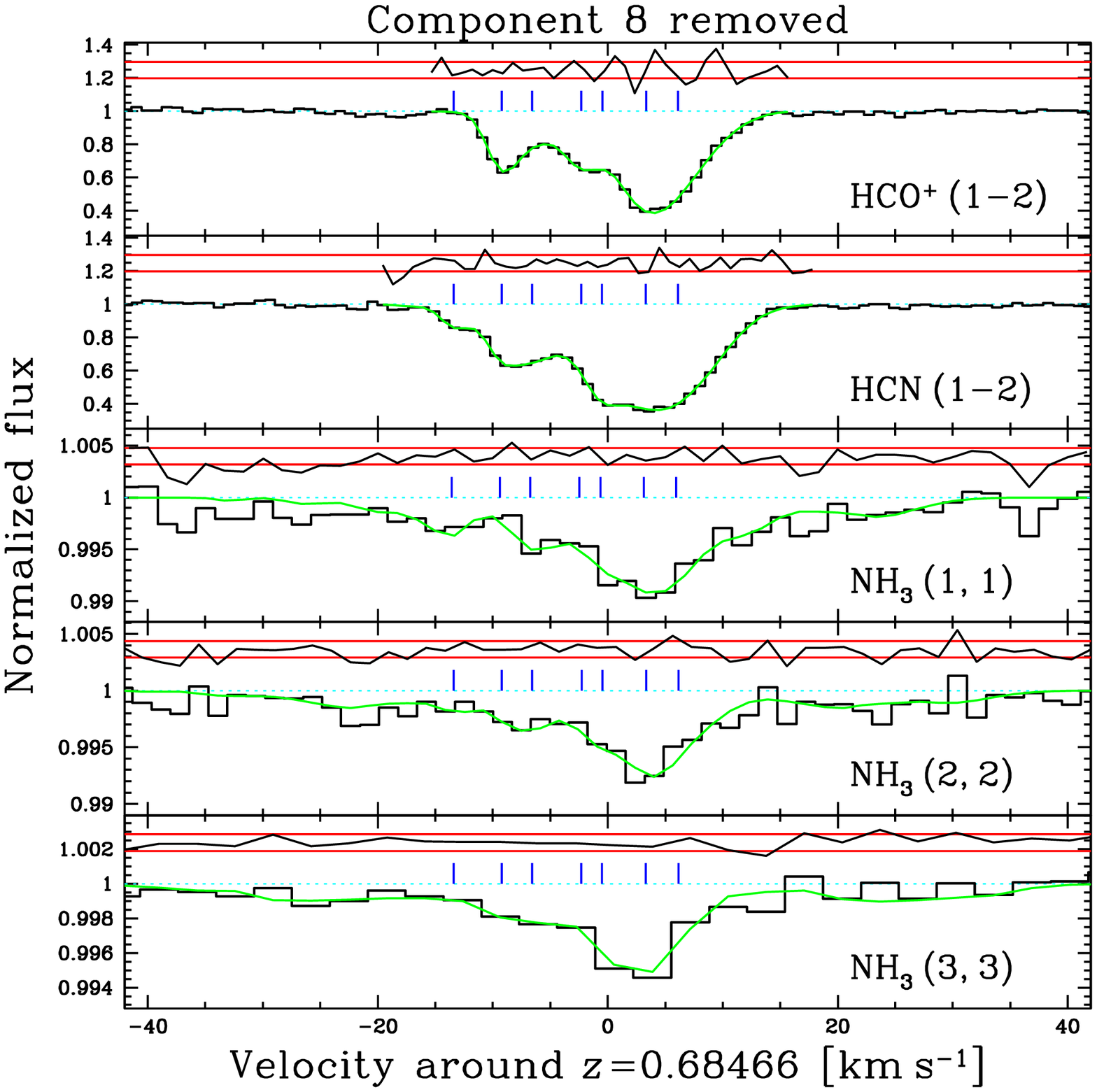}
   \hspace{0.01\textwidth}
   \includegraphics[width=0.32\textwidth]{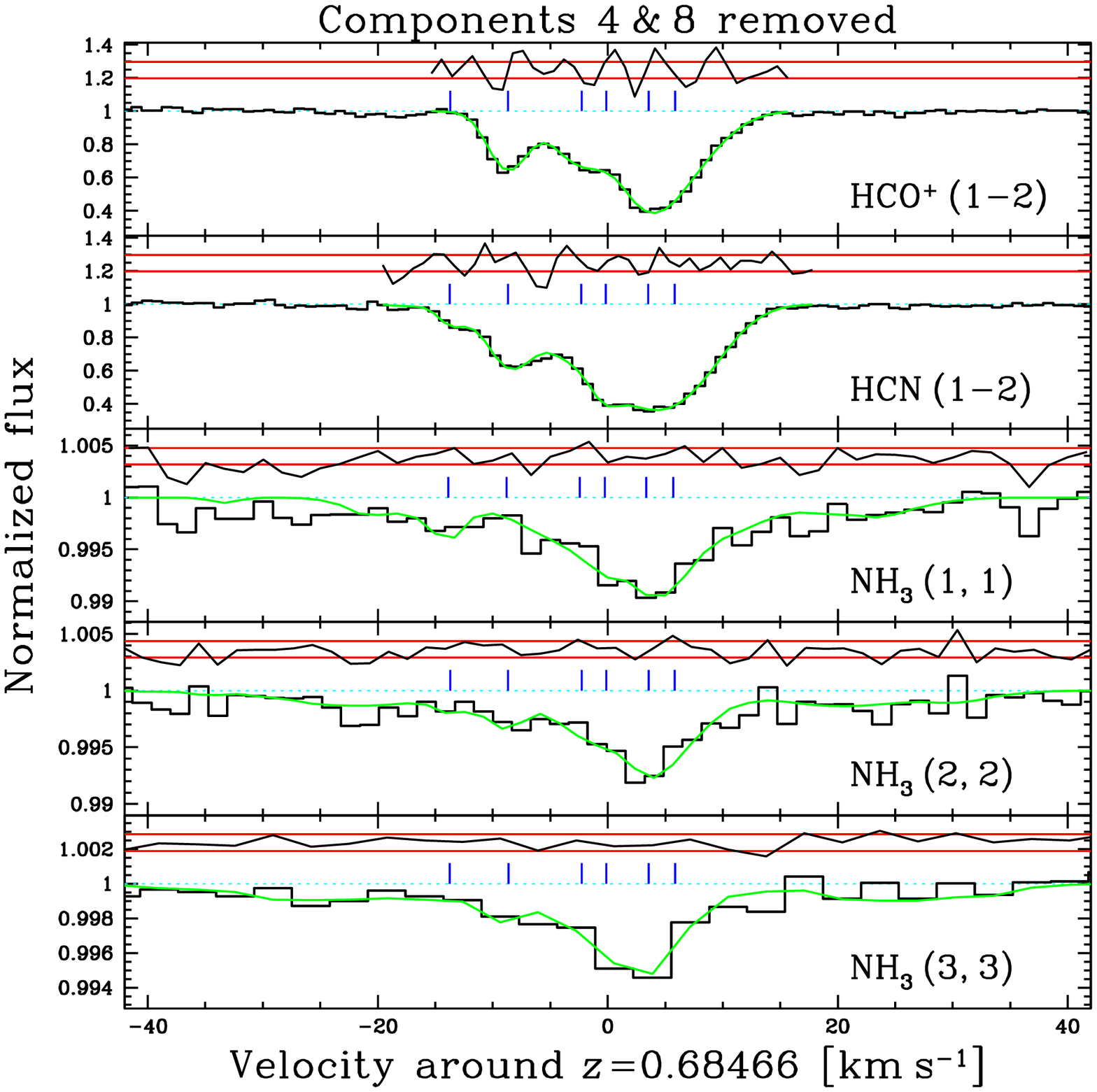}
   \hspace{0.01\textwidth}
   \includegraphics[width=0.32\textwidth]{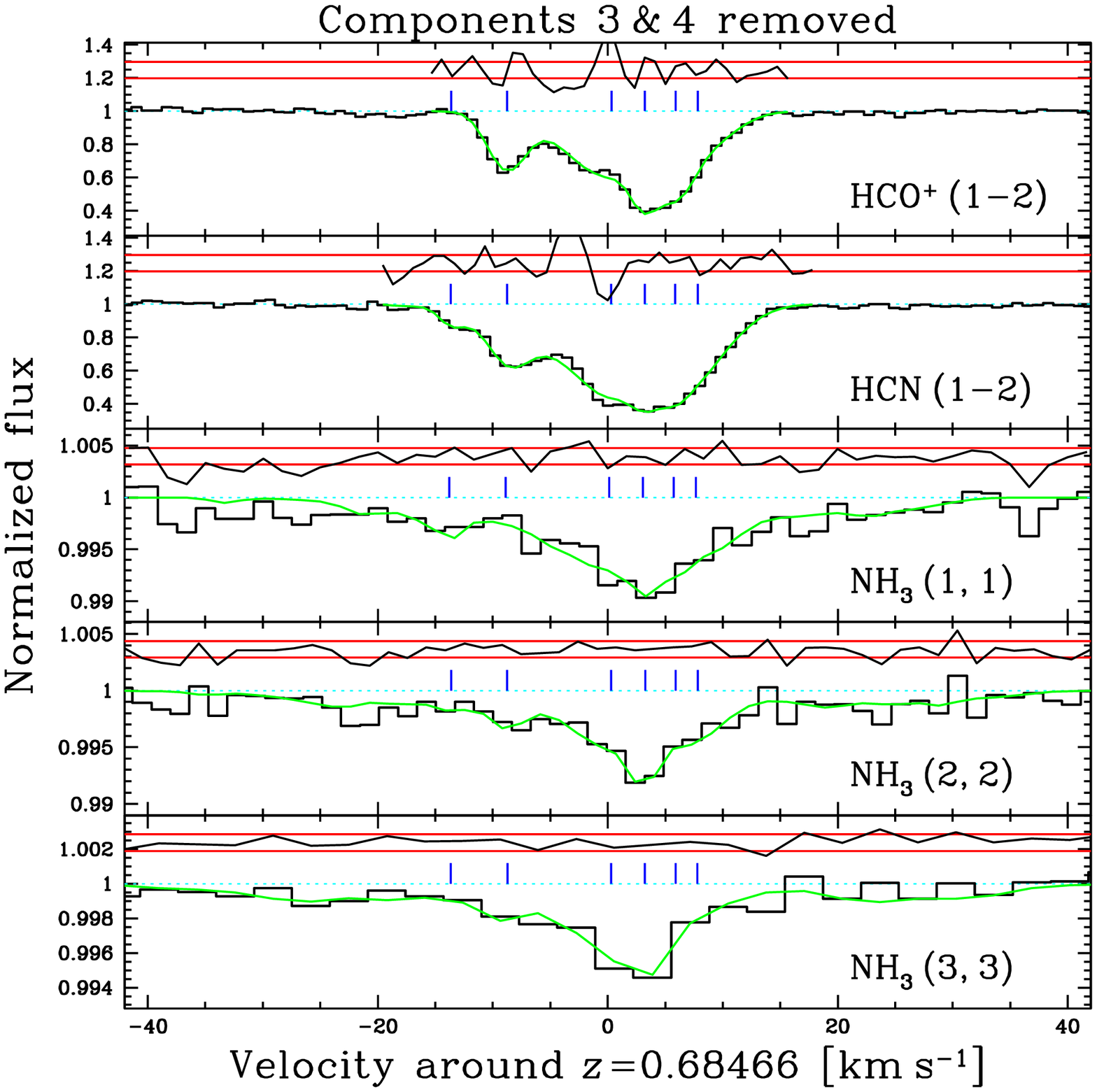}
  }
  \vspace{0.01\textheight}
  \hbox{
   \includegraphics[width=0.32\textwidth]{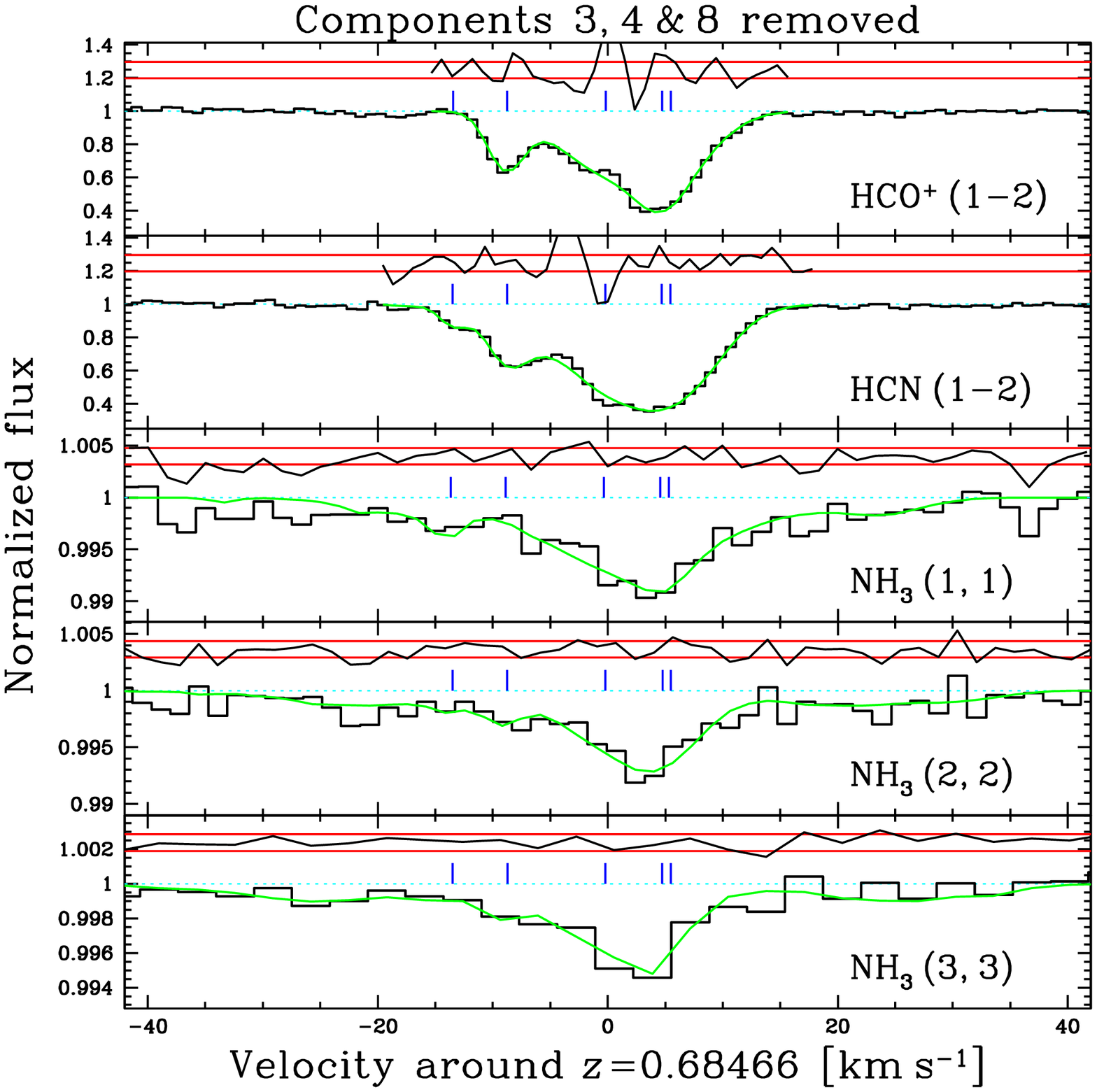}
   \hspace{0.01\textwidth}
   \includegraphics[width=0.32\textwidth]{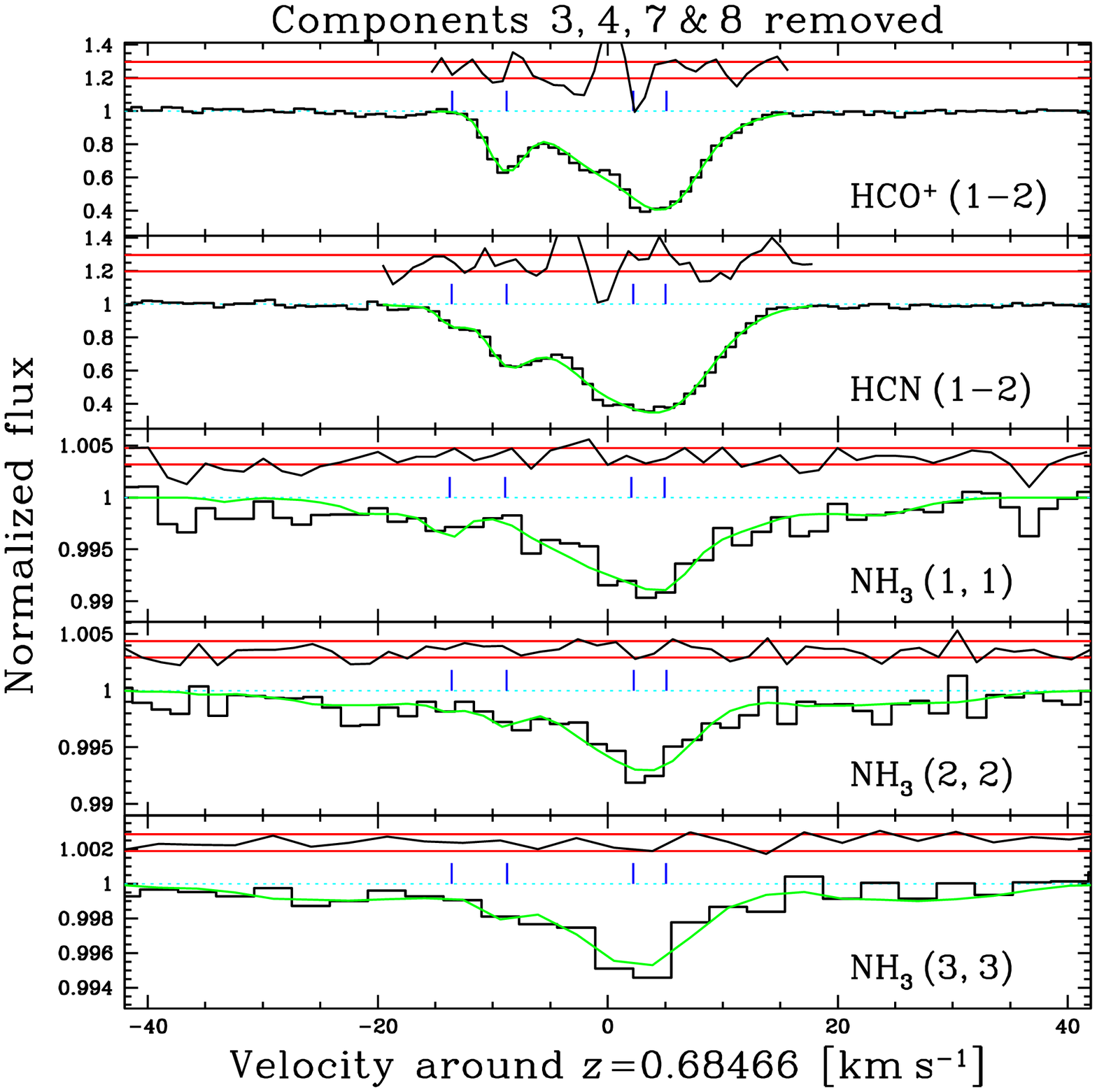}
  }
 }
}
\caption{The different velocity structures attempted in search of the
  statistically preferred one. The values of $\Delta\mu/\mu$ and
  $\chi^2_\nu$ corresponding to each fit are represented by the black
  points in Fig.~2. Components were removed/added from/to the fiducial
  fit to form each initial velocity structure and {\sc vpfit} was run
  again to minimize $\chi^2$. For the 9-component model, the
  additional component was added redwards of all other components. The
  residual spectra show how poor the fits are with $\le6$ components.}
\end{figure*}

\begin{figure}
\renewcommand \thefigure{S3}
\centerline{\includegraphics[width=\textwidth]{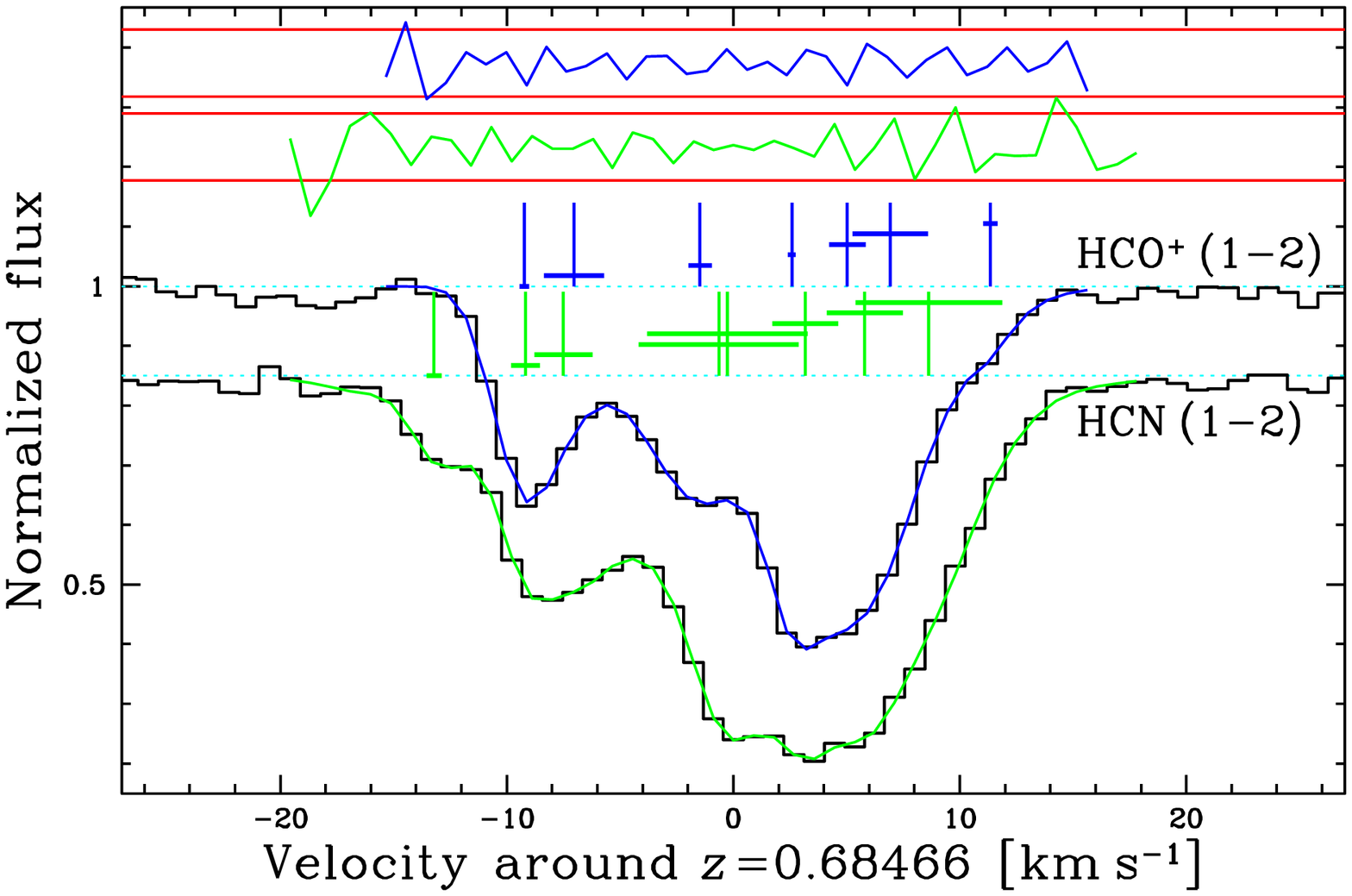}}
\caption{Independent fits to the two rotational transitions studied
  here. Layout similar to Fig.~1. The data, tick-marks and residuals
  for HCN(1--2), marked in lighter grey lines, are offset below those
  for HCO$^+$ for clarity. The formal 1-$\sigma$ statistical error in
  the redshift of each component (converted to velocity) is
  represented by the horizontal bars across the tick marks. The error
  bar for each component is plotted higher than the one to its left
  for clarity. There is broad agreement between the positions of the
  velocity components. The bluest component in HCN(1--2) is not
  statistically required by the HCO$^+$(1--2) data alone, similar to
  the simultaneous fit in Fig.~1.}
\end{figure}

\clearpage

\begin{table}
\renewcommand \thetable{S1}
\begin{center}
\caption{Laboratory frequencies of transitions used in our analysis.
  The second and third columns provide, respectively, the relevant
  quantum numbers for the lower and upper states in the following
  order: $J$, $F$, $F_1$. For HCO$^+$(1--2) only $J$ is relevant
  while $F_1$ is not relevant for HCN(1--2). For the NH$_3$
  inversion transitions, $J=K$, $F$ is the magnetic hyperfine
  quantum number and $F_1$ is the quadrupole quantum number. The
  final column shows the relative optical depth of different
  hyperfine components, $\Sigma$, in LTE. $\Sigma$ is normalized to
  unity for the strongest hyperfine component.}
\small
\begin{tabular}{lcclc}\hline
\multicolumn{1}{c}{Molecule}&\multicolumn{2}{c}{Transition}                     &\multicolumn{1}{c}{Frequency}&\multicolumn{1}{c}{$\log\Sigma$}\\
                            &\multicolumn{1}{c}{Lower}&\multicolumn{1}{c}{Upper}&\multicolumn{1}{c}{[GHz]}    &                                \\\hline
HCO$^+$                     & 1                       & 2                       & 178.375065(50)              & $\phantom{-}0.0000$            \\ \\
HCN                         & 1, 2                    & 2, 2                    & 177.2596770(20)             &           $-0.7482$            \\
HCN                         & 1, 0                    & 2, 1                    & 177.2599230(20)             &           $-0.6233$            \\
HCN                         & 1, 1                    & 2, 2                    & 177.2611100(20)             &           $-0.2711$            \\
HCN                         & 1, 2                    & 2, 3                    & 177.2612230(20)             & $\phantom{-}0.0000$            \\
HCN                         & 1, 2                    & 2, 1                    & 177.2620122(11)             &           $-1.9243$            \\
HCN                         & 1, 1                    & 2, 1                    & 177.2634450(20)             &           $-0.7482$            \smallskip\\
NH$_3$                      & 1, 1/2,  1              & 1, 1/2,  0              &  23.692926829(56)           &           $-0.7993$            \\
NH$_3$                      & 1, 3/2,  1              & 1, 1/2,  0              &  23.692968829(38)           &           $-0.4983$            \\
NH$_3$                      & 1, 1/2,  1              & 1, 3/2,  2              &  23.693872152(57)           &           $-0.7024$            \\
NH$_3$                      & 1, 3/2,  1              & 1, 5/2,  2              &  23.693905112(58)           &           $-0.4472$            \\
NH$_3$                      & 1, 3/2,  1              & 1, 3/2,  2              &  23.693914466(39)           &           $-1.4014$            \\
NH$_3$                      & 1, 1/2,  1              & 1, 1/2,  1              &  23.694459098(36)           &           $-1.1004$            \\
NH$_3$                      & 1, 1/2,  1              & 1, 3/2,  1              &  23.694470034(28)           &           $-1.4014$            \\
NH$_3$                      & 1, 3/2,  2              & 1, 5/2,  2              &  23.694470904(63)           &           $-1.1461$            \\
NH$_3$                      & 1, 3/2,  2              & 1, 3/2,  2              &  23.694480291(44)           &           $-0.1919$            \\
NH$_3$                      & 1, 3/2,  1              & 1, 1/2,  1              &  23.694501428(59)           &           $-1.4014$            \\
NH$_3$                      & 1, 5/2,  2              & 1, 5/2,  2              &  23.694505950(37)           & $\phantom{-}0.0000$            \\
NH$_3$                      & 1, 3/2,  1              & 1, 3/2,  1              &  23.694512322(37)           &           $-0.7024$            \\
NH$_3$                      & 1, 5/2,  2              & 1, 3/2,  2              &  23.694515319(47)           &           $-1.1461$            \\
NH$_3$                      & 1, 3/2,  2              & 1, 1/2,  1              &  23.695067195(46)           &           $-0.7024$            \\
NH$_3$                      & 1, 3/2,  2              & 1, 3/2,  1              &  23.695078206(59)           &           $-1.4014$            \\
NH$_3$                      & 1, 5/2,  2              & 1, 3/2,  1              &  23.695113176(50)           &           $-0.4472$            \\
NH$_3$                      & 1, 1/2,  0              & 1, 1/2,  1              &  23.696029719(48)           &           $-0.7993$            \\
NH$_3$                      & 1, 1/2,  0              & 1, 3/2,  1              &  23.696040646(38)           &           $-0.4983$            \smallskip\\
NH$_3$                      & 2, 3/2,  2              & 2, 3/2,  1              &  23.720534302(33)           &           $-1.7058$            \\
NH$_3$                      & 2, 5/2,  2              & 2, 3/2,  1              &  23.720575068(39)           &           $-0.7516$            \\
NH$_3$                      & 2, 3/2,  2              & 2, 1/2,  1              &  23.720579876(34)           &           $-1.0068$            \\
NH$_3$                      & 2, 5/2,  2              & 2, 7/2,  3              &  23.721336248(58)           &           $-0.7570$            \\
NH$_3$                      & 2, 3/2,  2              & 2, 5/2,  3              &  23.721337256(57)           &           $-0.9119$            \\
NH$_3$                      & 2, 5/2,  2              & 2, 5/2,  3              &  23.721377951(106)          &           $-2.0580$            \\
NH$_3$                      & 2, 1/2,  1              & 2, 3/2,  1              &  23.722588837(70)           &           $-1.0068$            
\end{tabular}
\normalsize
\end{center}
\end{table}

\begin{table}
\renewcommand \thetable{S1}
\begin{center}
\vskip\abovecaptionskip
\tablename~\thetable: {\it continued}
\vskip\belowcaptionskip
\small
\begin{tabular}{lcclc}\hline
\multicolumn{1}{c}{Molecule}&\multicolumn{2}{c}{Transition}                     &\multicolumn{1}{c}{Frequency}&\multicolumn{1}{c}{$\log\Sigma$}\\
                            &\multicolumn{1}{c}{Lower}&\multicolumn{1}{c}{Upper}&\multicolumn{1}{c}{[GHz]}    &                                \\\hline
NH$_3$                      & 2, 3/2,  2              & 2, 5/2,  2              &  23.722591529(106)          &           $-1.2621$            \\
NH$_3$                      & 2, 5/2,  3              & 2, 7/2,  3              &  23.722591879(86)           &           $-1.3010$            \\
NH$_3$                      & 2, 3/2,  2              & 2, 3/2,  2              &  23.722632304(33)           &           $-0.3079$            \\
NH$_3$                      & 2, 5/2,  3              & 2, 5/2,  3              &  23.722633644(33)           & $\phantom{-}0.0000$            \\
NH$_3$                      & 2, 1/2,  1              & 2, 1/2,  1              &  23.722634389(33)           &           $-0.7058$            \\
NH$_3$                      & 2, 5/2,  2              & 2, 3/2,  2              &  23.722673071(32)           &           $-1.2621$            \\
NH$_3$                      & 2, 7/2,  3              & 2, 5/2,  3              &  23.722675390(42)           &           $-1.3010$            \\
NH$_3$                      & 2, 3/2,  1              & 2, 1/2,  1              &  23.722679956(22)           &           $-1.0068$            \\
NH$_3$                      & 2, 5/2,  3              & 2, 5/2,  2              &  23.723887894(125)          &           $-2.0580$            \\
NH$_3$                      & 2, 5/2,  3              & 2, 3/2,  2              &  23.723928698(64)           &           $-0.9119$            \\
NH$_3$                      & 2, 7/2,  3              & 2, 5/2,  2              &  23.723929630(50)           &           $-0.7570$            \\
NH$_3$                      & 2, 1/2,  1              & 2, 3/2,  2              &  23.724686811(63)           &           $-1.0068$            \\
NH$_3$                      & 2, 3/2,  1              & 2, 5/2,  2              &  23.724691591(57)           &           $-0.7516$            \\
NH$_3$                      & 2, 3/2,  1              & 2, 3/2,  2              &  23.724732357(123)          &           $-1.7058$            \smallskip\\
NH$_3$                      & 3, 9/2,  3              & 3, 7/2,  2              &  23.867805094(22)           &           $-0.8013$            \\
NH$_3$                      & 3, 7/2,  3              & 3, 5/2,  2              &  23.867816691(45)           &           $-0.9651$            \\
NH$_3$                      & 3, 3/2,  3              & 3, 1/2,  2              &  23.867824516(112)          &           $-1.3541$            \\
NH$_3$                      & 3, 5/2,  3              & 3, 3/2,  2              &  23.867826808(20)           &           $-1.1500$            \\
NH$_3$                      & 3, 5/2,  3              & 3, 7/2,  4              &  23.868438244(47)           &           $-1.0669$            \\
NH$_3$                      & 3, 7/2,  3              & 3, 9/2,  4              &  23.868440344(24)           &           $-0.9428$            \\
NH$_3$                      & 3, 3/2,  3              & 3, 5/2,  4              &  23.868446261(71)           &           $-1.1938$            \\
NH$_3$                      & 3, 9/2,  3              & 3, 11/2, 4              &  23.868450126(71)           &           $-0.8258$            \\
NH$_3$                      & 3, 3/2,  2              & 3, 5/2,  2              &  23.870049079(21)           &           $-0.6059$            \\
NH$_3$                      & 3, 3/2,  3              & 3, 5/2,  3              &  23.870064875(46)           &           $-0.8734$            \\
NH$_3$                      & 3, 7/2,  3              & 3, 9/2,  3              &  23.870067403(53)           &           $-0.8557$            \\
NH$_3$                      & 3, 9/2,  4              & 3, 11/2, 4              &  23.870079000(37)           &           $-0.9130$            \\
NH$_3$                      & 3, 3/2,  3              & 3, 3/2,  3              &  23.870127881(20)           &           $-0.2713$            \\
NH$_3$                      & 3, 5/2,  4              & 3, 5/2,  4              &  23.870129616(20)           & $\phantom{-}0.0000$            \\
NH$_3$                      & 3, 1/2,  2              & 3, 1/2,  2              &  23.870130224(20)           &           $-0.7521$            \\
NH$_3$                      & 3, 11/2, 4              & 3, 9/2,  4              &  23.870180310(69)           &           $-0.9130$            \\
NH$_3$                      & 3, 9/2,  3              & 3, 7/2,  3              &  23.870188326(66)           &           $-0.8557$            \\
NH$_3$                      & 3, 5/2,  2              & 3, 3/2,  2              &  23.870211423(62)           &           $-0.6059$            \\
NH$_3$                      & 3, 11/2, 4              & 3, 9/2,  3              &  23.871807418(23)           &           $-0.8258$            \\
NH$_3$                      & 3, 5/2,  4              & 3, 3/2,  3              &  23.871811331(20)           &           $-1.1938$            \\
NH$_3$                      & 3, 9/2,  4              & 3, 7/2,  3              &  23.871817154(22)           &           $-0.9428$            \\
NH$_3$                      & 3, 7/2,  4              & 3, 5/2,  3              &  23.871819253(22)           &           $-1.0669$            \\
NH$_3$                      & 3, 3/2,  2              & 3, 5/2,  3              &  23.872431263(30)           &           $-1.1500$            \\
NH$_3$                      & 3, 1/2,  2              & 3, 3/2,  3              &  23.872433410(22)           &           $-1.3541$            \\
NH$_3$                      & 3, 5/2,  2              & 3, 7/2,  3              &  23.872441474(22)           &           $-0.9651$            \\
NH$_3$                      & 3, 7/2,  2              & 3, 9/2,  3              &  23.872452975(39)           &           $-0.8013$            \\\hline
\end{tabular}
\normalsize
\end{center}
\end{table}

\begin{table}
\begin{center}
\renewcommand \thetable{S2}
\caption{Absorption line parameters and 1-$\sigma$ uncertainties for 
  the 8-component fiducial fit in Fig.~1. The optical depths, $\tau$,
  for the HCN and NH$_3$ transitions are summed over all hyperfine
  components. For NH$_3$, $\tau$ is given in units of 0.01.}
\small
\begin{tabular}{cccccccc}\hline
Com-      & Redshift       & FWHM           & \multicolumn{2}{c}{$\tau$}                            & \multicolumn{3}{c}{$\tau/10^{-2}$}                                      \\
ponent    &                & [km\,s$^{-1}$] & HCO$^+$(1--2)             & HCN(1--2)                 & NH$_3$(1,1)            & NH$_3$(2,2)            & NH$_3$(3,3)           \\\hline
1         & 0.6845833(12)  & $2.80\pm0.40$  & $0.004^{+0.011}_{-0.003}$ & $0.152^{+0.016}_{-0.014}$ & $0.31^{+0.25}_{-0.14}$ & $0.07^{+0.50}_{-0.06}$ & $0.19^{+0.28}_{-0.10}$\\
2         & 0.6846071(12)  & $3.00\pm0.20$  & $0.41^{+0.12}_{-0.10}$    & $0.45^{+0.24}_{-0.16}$    & $0.00^{+1.3}_{-1.1}$   & $0.37^{+0.42}_{-0.20}$ & $0.36^{+0.42}_{-0.20}$\\
3         & 0.6846205(34)  & $3.56\pm1.07$  & $0.15^{+0.11}_{-0.06}$    & $0.31^{+0.16}_{-0.11}$    & $0.61^{+0.28}_{-0.19}$ & $0.32^{+0.31}_{-0.16}$ & $0.23^{+0.32}_{-0.13}$\\
4         & 0.6846473(133) & $5.11\pm2.53$  & $0.39^{+0.68}_{-0.25}$    & $0.33^{+0.58}_{-0.21}$    & $0.44^{+0.91}_{-0.30}$ & $0.57^{+1.12}_{-0.38}$ & $0.40^{+0.89}_{-0.28}$\\
5         & 0.6846568(9)   & $3.14\pm0.37$  & $0.11^{+0.99}_{-0.10}$    & $0.74^{+0.46}_{-0.28}$    & $0.84^{+0.71}_{-0.38}$ & $0.37^{+7.10}_{-0.35}$ & $0.37^{+2.74}_{-0.32}$\\
6         & 0.6846730(19)  & $2.86\pm0.98$  & $0.59^{+1.18}_{-0.40}$    & $0.60^{+1.10}_{-0.39}$    & $1.26^{+1.30}_{-0.64}$ & $1.14^{+1.50}_{-0.65}$ & $0.85^{+1.08}_{-0.48}$\\
7         & 0.6846879(38)  & $4.25\pm1.27$  & $0.63^{+0.91}_{-0.37}$    & $0.72^{+1.38}_{-0.47}$    & $0.86^{+2.00}_{-0.60}$ & $0.32^{+4.78}_{-0.30}$ & $0.18^{+5.80}_{-0.18}$\\
8         & 0.6847024(138) & $6.55\pm2.01$  & $0.25^{+0.42}_{-0.16}$    & $0.40^{+0.62}_{-0.24}$    & $0.43^{+0.87}_{-0.29}$ & $0.46^{+0.70}_{-0.28}$ & $0.20^{+0.44}_{-0.14}$\\\hline
\end{tabular}
\normalsize
\end{center}
\end{table}


\begin{thebibliography}{10}
\bibitem{cosmology}
Throughout we employ a standard $\Lambda$ cosmology with $H_0=71{\rm \,km\,s}^{-1}{\rm \,Mpc}^{-1}$, $\Omega_{\rm m}=0.27$ and $\Omega_\Lambda=0.73$ whereby $z=0.68466$ corresponds to a look-back time of 6.2\,Gyr ($\sim$half the age of the Universe) with a linear angular scale of $7.1{\rm \,pc\,mas}^{-1}$.
\end{thebibliography}
\end{document}